\documentclass[twoside]{article}

\usepackage{lipsum} 

\usepackage[sc]{mathpazo} 
\usepackage[T1]{fontenc} 
\usepackage{microtype} 

\usepackage[hmarginratio=1:1,top=32mm,columnsep=20pt]{geometry} 
\usepackage{multicol} 
\usepackage[hang, small,labelfont=bf,up,textfont=it,up]{caption} 
\usepackage{booktabs} 
\usepackage{float} 
\usepackage{hyperref} 
\hypersetup{%
    pdfborder = {0 0 0}
}
\usepackage[section]{placeins}

\usepackage{lettrine} 
\usepackage{paralist} 

\usepackage{setspace}
\usepackage{amssymb,latexsym}
\usepackage{amsmath}
\usepackage{ifpdf}
\usepackage[pdftex]{graphicx}
\usepackage{amstext}
\usepackage{amssymb}
\usepackage{tocloft}

\usepackage{textcomp}
\usepackage{booktabs}

\usepackage{enumerate}
\usepackage{mathtools}
\usepackage{units}

\usepackage{bm}
\usepackage{amsbsy}
\usepackage{chapterbib}
\usepackage{cite} 
\usepackage{authblk}
\usepackage[ruled,vlined]{algorithm2e}

\usepackage{abstract} 

\usepackage{titlesec} 
\renewcommand\thesection{\arabic{section}} 
\renewcommand\thesubsection{\arabic{subsection}} 
\titleformat{\section}[block]{\large\scshape\centering}{\thesection.}{1em}{} 
\titleformat{\subsection}[block]{\large}{\thesubsection.}{1em}{} 

\usepackage{fancyhdr} 
\title{\vspace{-15mm}\fontsize{24pt}{10pt}\selectfont\textbf{A Dynamic Approach to Linear Statistical Calibration with an Application in Microwave Radiometry}} 

\author[]{Derick L. Rivers\thanks{Corresponding author. Email:~riversdl@vcu.edu} ~and Edward L. Boone}
\affil{Department of Statistical Sciences and Operations Research, Virginia Commonwealth University}
\date{}

\begin{document}

\maketitle 

\thispagestyle{fancy} 


\begin{abstract}

The problem of statistical calibration of a measuring instrument can be framed both in a statistical context as well as in an engineering context. In the first, the problem is dealt with by distinguishing between the ``classical" approach and the ``inverse" regression approach. Both of these models are static models and are used to estimate ``exact" measurements from measurements that are affected by error. In the engineering context, the variables of interest are considered to be taken at the time at which you observe it. The Bayesian time series analysis method of Dynamic Linear Models (DLM) can be used to monitor the evolution of the measures, thus introducing an {\it dynamic} approach to statistical calibration. The research presented employs the use of Bayesian methodology to perform statistical calibration. The DLM's framework is used to capture the time-varying parameters that maybe changing or drifting over time. Two separate DLM based models are presented in this paper.  A simulation study is conducted where the two models are compared to some well known 'static' calibration approaches in the literature from both the frequentist and Bayesian perspectives. The focus of the study is to understand how well the {\it dynamic statistical calibration} methods performs under various signal-to-noise ratios, $r$. The posterior distributions of the estimated calibration points  as well as the $95\%$ coverage intervals are compared by statistical summaries. These dynamic methods are applied to a microwave radiometry dataset.\\

\end{abstract}



\section{Introduction}\label{sec:Introduction}

Calibrating measurement instruments is a important problem that engineers frequently need to address. There exist several statistical methods that address this problem that are based on a simple linear regression approach. In tradition simple linear regression the goal is to relate a known value of X to a uncertain value of Y using a linear relationship. In contrast, the statistical calibration problem seeks to utilize a simple linear regression model to relate a known value of Y to an uncertain value of X. This is why statistical calibration is sometimes called {\it inverse regression} due to its relationship to simple linear regression (Osborne 1991; Ott and Longnecker 2009). Recall in linear regression the model is given as follows:
\begin{equation}\label{eq:linear_matrix_error}
{\bf Y} = {\bf X} {\boldsymbol \beta} + {\boldsymbol \epsilon}
\end{equation}
 where {\bf Y} is a $(n \times 1)$ response vector, {\bf X} is a $(n \times p)$ matrix of independent variables with $p=k+1$ total model parameters, ${\boldsymbol \beta}$ is a $(p \times 1)$ vector of unknown fixed parameters and ${\boldsymbol \epsilon}$ is a $(n \times 1)$ vector of uncorrelated error terms with zero mean (Myers 1990; Draper and Smith 1998; Montgomery {\it et al.} 2012). It is assumed that the value of the predictor variable {\bf X} = {\bf x} are nonrandom and observed with negligible error, while the $n$ error terms are random variables with mean zero and constant variance $\sigma^{2}$  (Myers 1990).
Typically, in regression, of interest is the estimation of the parameter vector; ${\boldsymbol \beta}$, and possibly the prediction of a future value $\hat{\bf Y}_{i|new}$ corresponding to a new ${\bf X} = x_{i|new}^{\prime}$ value. The prediction problem is relatively straightforward, due to the fact that a future ${\bf Y}_{i}$ value can be made directly by substituting $ x_{i|new}^{\prime}$ into (\ref{eq:linear_matrix_error}) with $E[\epsilon]=0$.\\
\indent For the statistical calibration problem let $y_{0}$ be the known {\it observed} value of the response and $x_{0}$ be the corresponding regressor, $x_{0}$ which is to be estimated. This problem is conducted in two stages: first measurement pairs $(x_{i}, y_{i})$ of data is observed and a simple linear regression line is fit by estimating ${\boldsymbol \beta}$; secondly, $m$ observations of the response are observed, all corresponding to a single $x_{0}$ (\"{O}zyurt and Erar 2003). Since $y_{0}$ is fixed, inferences are different than those in a traditional regression (or prediction) problem (Osborne 1991; Eno 1999; Eno and Ye 2000).

		\subsection{Classical Calibration Methods}
\indent Eisenhart (1939) offered the first solution to the calibration problem, and is commonly known as the $``classical"$ estimator to the linear calibration problem. They assumed that the relationship between $x$ and $y$ was of a simple linear form:
\begin{equation*}
E(Y|X = x) = \beta_{0} + \beta_{1} x.
\end{equation*}
The estimated regression line for the first stage of the experiment is given by
\begin{equation}\label{eq:simple}
\hat{Y} = \hat{\beta_{0}} + \hat{\beta}_{1} X, 
\end{equation}
where $\hat{\beta_{0}}$ and $\hat{\beta_{1}}$ are the least squares estimate of $\beta_{0}$ and $\beta_{1}$, respectively.
~Using the data collected at the first stage of experimentation, Eisenhart (1939) inverts Equation (\ref{eq:simple}) to estimate the unknown regressor value $x_{0}$ for an observed response value $y_{0}$, by:
\begin{equation}\label{eq:ClassEq}
\hat{x}_{0,c} = \frac{y_{0} - \hat{\beta}_{0}}{\hat{\beta}_{1}}
\end{equation}
where $\hat{x}_{0,c}$ denotes the $``classical"$ estimator for $x_{0}$. Since division by $\hat{\beta}_{1}$ is used there is an implicit assumption that $\lvert\hat{\beta}_{1}\rvert>0$.\\ 
\indent Assuming that $\lvert\hat{\beta}_{1}\rvert>0$, Brown (1993) describes the following interval estimate corresponding to Eisenhart (1939): 
\begin{equation*}\label{eq:ClassInv}
\frac{y_{0} - \hat{\beta}_{0}}{\hat{\beta}_{1}} \left(1 + \frac{\hat{\sigma}^{2} t^{2}}{\hat{\beta}_{1}^{2} S_{xx}}\right) \pm \frac{\hat{\sigma}t}{\hat{\beta}_{1}} \left(1+\frac{1}{2n}+\frac{(y_{0} - \hat{\beta}_{0})^{2}+\hat{\sigma}^2t^{2}}{2\hat{\beta}_{1}^{2}S_{xx}}\right),
\end{equation*}
where 
\begin{equation}
\hat{\sigma} = \sqrt{\frac{\sum_{i=1}^{n}(y_{i}-\hat{\beta}_{0}-\hat{\beta}_{1}x_{i})^{2}}{n-2}},\nonumber
\end{equation}
\begin{equation}
S_{xx} = \sum_{i=1}^{n}(x-\bar{x})^{2},\nonumber
\end{equation}
\indent Krutchkoff (1967) proposed a competitive approach to Eisenhart's (1939) classical linear calibration solution, which he called the $``inverse"$ regression calibration method and is written as:
\begin{equation}
X_{i} = \phi + \delta Y_{i} + \epsilon^{'}_{i},\nonumber
\end{equation}
where $\phi$ and $\delta$ are the parameters in the linear relationship and $\epsilon^{'}_{i}$ are independent identically distributed measurement errors with a zero mean and finite variance. Here $\phi$ and $\delta$ are estimated via least squares. The unknown $x_{0}$ can be estimated directly by substituting $y_{0}$ into the fitted equation: 
\begin{equation}\label{eq:InvEq}
\hat{x}_{0,I} = \hat{\phi} + \hat{\delta} y_{0}.
\end{equation}
We let $\hat{x}_{0,I}$ denote the $``inverse"$ estimator of $x_{0}$. The $100(1-\alpha)\%$ confidence interval for $E(x_{0,I}|y_{0})$ can be written as
\begin{equation*}
x_{0,I}(y_{0}) \pm t_{\nicefrac{\alpha}{2}}\hat{\sigma}\sqrt{\frac{1}{n}+\frac{(y_{0}-\bar{y})^{2}}{S_{yy}}}
\end{equation*}
where
\begin{equation*}
S_{yy} = \sum_{i=1}^{n}(y_{i}-\bar{y})^{2}.
\end{equation*}
\indent Krutchkoff (1967) used a simulation study, where he found that the mean squared error of estimation for $x_{0}$ was uniformly less for this estimator versus the classical estimator. The inverse approach was later supported by Lwin and Maritz (1982). For criticisms of Krutchkoff's (1967) approach such as bias see Osborne (1991).
		\subsection{Bayesian Calibration Methods}
\indent The first noted Bayesian solution to the calibration problem was presented by Hoadley (1970). His work was motivated by the unanswered question in the Frequentist community of whether $\beta_{1}$ is zero (or close to zero). Hoadley (1970) justified the use of the $``inverse"$ estimator (Krutchkoff, 1967) by considering the ususal $F$-statistic to test the hypothesis that $\beta_{1} = 0$ where $F = \hat{\beta}_{1}^{2}S_{xx}/\hat{\sigma}^{2}$,
\begin{equation*}
\hat{\sigma}^{2} = \frac{ \left\{ \sum_{i=1}^{n}\left(y_{1i} - (\hat{\beta}_{0} + \hat{\beta}_{1} x_{i})\right)^{2} + \sum_{j=1}^{m} \left( y_{2j} - \bar{y}_{2}\right)^{2}  \right\} }{ (n+m-3)}.
\end{equation*}
\indent The assumption made by Hoadley (1970) reflects that $x_{0}$ is random and {\it a priori} independent of $\pi(\beta_{0}, \beta_{1}, \sigma^{2})$, so that the joint prior distribution of $\pi(\beta_{0}, \beta_{1}, \sigma^{2}, x_{0}) \propto \pi(\beta_{0}, \beta_{1}, \sigma^{2}) \pi(x_{0})$. Hoadley (1970) first assumed that $(\beta_{0}, \beta_{1}, \sigma^{2})$ had a uniform distribution,
\begin{equation}
\pi(\beta_{0}, \beta_{1}, \sigma^{2}) \propto \sigma^{-2},\nonumber
\end{equation}
 but the prior distribution for $x_{0}$ was not given.\\ 
\indent Hoadley (1970) shows for $m = 1$ (one observation at the prediction stage), that if $x_{0}$ has a prior density from a Student {\it t} distribution with $n-3$ degrees of freedom, a mean of 0, and a scale parameter  
\begin{equation*}
\sigma = \frac{n+1}{n-3},
\end{equation*}
the posterior distribution is
\begin{equation}\label{eq:HoadEq}
\pi(x_{0}|{\bf Data}) = t_{n-2}\left(\hat{x}_{0,I}, \left[\frac{n+1+ (\hat{x}_{0,I})^{2}/R}{F+n-2}\right]\right),
\end{equation}
where $\hat{x}_{0,I}$ is the inverse estimator given by (\ref{eq:InvEq})$, R = \frac{F}{F+n-2}$ and $F = \hat{\beta}_{1}^{2}S_{xx}/\hat{\sigma}^{2}$.\\ 
\indent Hunter and Lamboy (1981) also considered the calibration problem from a Bayesian point of view and is similar to that of Hoadley (1970) because both assume the prior distribution to be
\begin{equation*}
\pi(\beta_{0}, \beta_{1}, \sigma^{2}, \eta) \propto \sigma^{-2}
\end{equation*}
where $\eta = \beta_{0} + \beta_{1}x_{0}$ which is the predicted $y_{0}$. The primary difference between their approach and the approach of Hoadley (1970) is that {\it a priori} they assume that $\eta$ and $(\beta_{0}, \beta_{1}, \sigma^{2})$ are independent while Hoadley (1970) assumed {\it a priori} that $x_{0}$ and $(\beta_{0}, \beta_{1}, \sigma^{2})$ are independent.\\
\indent Hunter and Lamboy (1981) uses an approximation to the posterior distribution of the unknown regressor $x_{0}$ by
\begin{equation}\label{eq:HuntEq}
	\pi(x_{0}|{\bf Data}) = N\left(\hat{x}_{0,c}, \frac{(s_{11} + s_{33})s_{22}- s_{12}^{2}}{s_{22}\hat{\beta}_{1}^{2}}\right),
\end{equation}
where
\[ {\bf S} = \{s_{i,j}\} = 
\left[ \begin{array}{ccc}
s_{11}  & s_{12}  & 0 \\
s_{12}  & s_{22}  & 0 \\
0          & 0           & s_{33} \end{array} \right] =
\left[ \begin{array}{cc}
({\bf X}^{'} {\bf X})^{-1}\hat{\sigma}^{2}  & {\bf 0}  \\
{\bf 0}               &\nicefrac{ \hat{\sigma}^{2}}{m}   \end{array} \right],\] 
with $\hat{x}_{0,c}$ being the classical estimator given in Equation (\ref{eq:ClassEq}), $s_{i,j}$ denote the element of the $i^{th}$ row and $j^{th}$ column from variance-covariance matrix of the joint posterior density of ($\beta_{0}$, $\beta_{1}$, $\eta$).\\
\indent The remainder of this paper is organized as follows. Section \ref{sec:Dyn_Cal} presents the development of the dynamic approaches to the statistical calibration problem. In Section \ref{sec:Sim_Study} the results from the simulation study where the dynamics methods are evaluated along with the static approaches are presented. In Section \ref{sec:application} the proposed methods are applied to microwave radiometer data. In Section \ref{sec:future} future work and other considerations are given.

\section{Dynamic Calibration Approach}\label{sec:Dyn_Cal}

Traditional calibration methods assume the regression relationship is ``static'' in time. In many cases this is false, for example in microwave radiometry the static nature of the relationship is known to change across time. A dynamic approach can be created by letting the regression coefficients vary through time,
\begin{equation*}\label{eq:simp1time}
y_{t} = \beta_{0t} + \beta_{1t}x_{t} + \epsilon_{t},
\end{equation*}
where $\epsilon_{t} \stackrel{iid}{\sim} N[0, \sigma^{2}_{t}]$ and is known as the $observational$ error.\\
%
\indent The model may have different defining parameters at different times. One approach is to model $\beta_{0t}$ and $\beta_{1t}$ by using random walk type evolutions for the defining parameters, such as:
\begin{eqnarray*}
\beta_{0t} &=& \beta_{0(t-1)} + \omega_{\beta_{0t}}, \label{eq:paramevol1} \\
\beta_{1t} &=& \beta_{1(t-1)} + \omega_{\beta_{1t}}, \label{eq:paramevol2} 
\end{eqnarray*}
where $\omega_{\beta_{0t}}$ and $\omega_{\beta_{1t}}$ are independent zero-mean error terms with finite variances.
At any time $t$ the calibration problem is given by:
\begin{equation*}
y_{0t} = \beta_{0t} + \beta_{1t}x_{0t} + \epsilon_{t},  \hspace{30pt} t = 1,2,\dots,T. \label{eq:dynamic_x}
\end{equation*}
Bayesian Dynamic Linear Models (DLMs) approach of West {\it et al.} (1985); West and Harrison (1997) can be employed to achieve this goal. Recall the DLM framework is:
\begin{eqnarray*}
\mbox{Observation equation}: \hspace{1cm} & {\bf Y}_{t} = {\bf X}_{t} {\bf X}_{t} {\boldsymbol \theta_{t}} + {\boldsymbol \epsilon}_{t},  \hspace{0.9cm} & {\boldsymbol \epsilon}_{t} \sim N_{r}[{\bf 0,  E}]\label{eq:observe}\\
\mbox{System equation}: \hspace{1cm} &  {\boldsymbol \theta_{t}} = {\bf G}_{t}{\boldsymbol \theta_{t-1}} + {\boldsymbol \omega_{t}},  & {\boldsymbol \omega}_{t} \sim N_{d}[{\bf 0, W}]\label{eq:system}\\
\mbox{Initial information}: \hspace{1cm} & ({\boldsymbol \theta_{0}} | D_{0}) \sim  N_{d}[{\bf m_{0}, C_{0}}],
\end{eqnarray*} 
for some prior mean {$\bf m_{0}$} and variance {$\bf C_{0}$} with the vector of error terms, ${\boldsymbol \epsilon}_{t}$ and ${\boldsymbol \omega}_{t}$ independent across time and at any time.\\
\indent To update the model through time West and Harrison (1997) give the following method:
	\begin{enumerate}[(a)]
\item Posterior distribution at $t-1$: For some mean ${\bf m}_{t-1}$ and variance ${\bf C}_{t-1}$,\\
		\mbox{\hspace{90pt}} $({\boldsymbol \theta_{t-1}} | D_{t-1}) \sim N_{d}[{\bf m}_{t-1}, {\bf C}_{t-1}]$.
\item Prior distribution at time $t$: $({\boldsymbol \theta_{t}} | D_{t-1}) \sim N_{d}[{\bf a}_{t}, {\bf R}_{t}]$, where\\
		\mbox{\hspace{90pt}} ${\bf a}_{t} = {\bf G}_{t}{\bf m}_{t-1}$ \mbox{\hspace{20pt}} and \mbox{\hspace{20pt}} ${\bf R}_{t} = {\bf G}_{t}{\bf C}_{t-1}{\bf G}^{'}_{t} + {\bf W}$.
\item One-step forecast: $({\bf Y}_{t} | D_{t-1}) \sim N_{r}[{\bf f}_{t}, {\bf Q}_{t}]$, where\\
		\mbox{\hspace{90pt}} ${\bf f}_{t} = {\bf X}_{t} {\bf a}_{t}$ \mbox{\hspace{20pt}} and \mbox{\hspace{20pt}} ${\bf Q}_{t} = {\bf X}_{t} {\bf R}_{t}{\bf X}_{t} ^{'} + {\bf E}$.
\item Posterior distribution at time $t$: $({\boldsymbol \theta_{t}} | D_{t}) \sim N_{d}[{\bf m}_{t}, {\bf C}_{t}]$, with\\
		\mbox{\hspace{90pt}} ${\bf m}_{t} = {\bf a}_{t} + {\bf A}_{t}{\bf e}_{t}$ \mbox{\hspace{20pt}} and \mbox{\hspace{20pt}} ${\bf C}_{t} = {\bf R}_{t} - {\bf A}^{'}_{t}{\bf Q}_{t}{\bf A}_{t},$\\
		where\\
\mbox{\hspace{90pt}} ${\bf A}_{t} = {\bf Q}^{-1}_{t}{\bf X}_{t} {\bf R}_{t}$ \mbox{\hspace{20pt}} and \mbox{\hspace{20pt}} ${\bf e}_{t} = {\bf Y}_{t} - {\bf f}_{t}$.
	\end{enumerate}

\indent The DLM framework is used to establish the evolving relationship between the fixed design matrix ${\bf X}_{t}$  and ${\bf Y}_{t}$ by estimating ${\boldsymbol \theta}_{t}$, which is a $(d \times n)$ matrix of time-varying regression coefficients $\beta_{0t}$ and $\beta_{1t}$. For our calibration situation ${\bf Y}_{t}$ is a $(r \times n)$ matrix of responses and ${\bf G}_{t}$ is a known ($d \times d$) system matrix. The error ${\boldsymbol \epsilon}_{t}$ and ${\boldsymbol \omega}_{t}$ are independent normally distributed random $(r \times n)$ matrices with zero mean and constant variance-covariance matrices {\bf E} and {\bf W}. For simplification ${\bf G}_{t}$ is set equal to ${\bf I}_{(d \times d)}$, ${\bf E}$ is set equal to $\sigma^{2}_{E}{\bf I}_{(r \times r)}$ and ${\bf W}$ is $\sigma^{2}_{W}\left[{\bf X}_{t} ^{'} {\bf X}_{t} \right]^{-1}$. The past information is contained in the set $D_{0}$.\\ 
\indent  We specify a prior in the first stage of calibration for the unknown variances and derive an algorithm to draw from the posterior distribution of the unknown parameters,
\begin{equation*}
\pi({\boldsymbol \theta}_{t}, \sigma^{2}_{E}, \sigma^{2}_{W}|{\bf Y}_{t}) \propto \pi({\boldsymbol \theta}_{t}|\sigma^{2}_{E}, \sigma^{2}_{W},{\bf Y}_{t})\pi(\sigma^{2}_{E}, \sigma^{2}_{W}|{\bf Y}_{t})\label{eq:fulljoint}.
\end{equation*}
The second stage of the calibration experiment consists of using the joint posterior distribution $\pi({\boldsymbol \theta}_{t},\sigma^{2}_{E}, \sigma^{2}_{W}|{\bf Y}_{t})$ to derive $x_{0t}| {\boldsymbol \theta}_{t},\sigma^{2}_{E}, \sigma^{2}_{W}$ for each draw of $\pi(\sigma^{2}_{E}, \sigma^{2}_{W}|{\bf Y}_{t})$. The estimator for the parameter of interest, $x_{0t}$, is defined in a manner akin to Eisenhart (1939); Hunter and Lamboy (1981); Eno (1999), where
\begin{equation}\label{eq:dlm_calib1}
x_{0t} = \frac{y_{0t} - \beta_{0t}}{\beta_{1t}}.
\end{equation}
\indent In the final stage of the calibration experiment, the posterior distribution summary statistics are gathered at each time point $t$. The posterior median and credible intervals are taken for each $t$ across the draws of $x_{0t}| {\boldsymbol \theta}_{t},\sigma^{2}_{E}, \sigma^{2}_{W}$. The result of the dynamic calibration experiment is a time series of calibration distributions across time. We will be able to observe the distributional changes of the system with respect to the calibration reference.\\
\indent The proposed calibration estimator is developed by first considering the joint posterior distribution $\pi(\sigma^{2}_{E}, \sigma^{2}_{W}|{\bf Y}_{t})$.
We let ${\boldsymbol \Gamma}$ denote the vector of unknown DLM dispersion parameters where ${\boldsymbol \Gamma}^{\prime} = (\sigma^{2}_{E}, \sigma^{2}_{W})$. The prior information for the dispersion parameters is described by a prior density $\pi({\boldsymbol \Gamma})$ which summarizes what is known about the variance parameters before any data are observed. Using the Bayesian inferential approach, the prior information about the parameters must be combined with information contained in the data. The information provided by the data is captured by the likelihood functions, $f_{{\bf Y}}({\bf Y}_{t}|{\boldsymbol \theta}_{t}, \sigma^{2}_{E}, \sigma^{2}_{W})$ and $f_{\boldsymbol \theta}({\boldsymbol \theta}_{t}|{\boldsymbol \theta}_{t-1}, \sigma^{2}_{W})$ for the observation equation and the system equation, respectively. The combined information is described by the posterior density using the Bayes theorem (Bernardo and Smith 1994) as
\begin{equation*}
\pi({\boldsymbol \Gamma}|{\bf Y}_{t}) \propto f_{{\bf Y}}({\bf Y}_{t}|{\boldsymbol \theta}_{t}, \sigma^{2}_{E}, \sigma^{2}_{W}) \cdot f_{\boldsymbol \theta}({\boldsymbol \theta}_{t}|{\boldsymbol \theta}_{t-1}, \sigma^{2}_{W})\cdot \pi({\boldsymbol \Gamma}).
\end{equation*}
\indent For our calibration problem it is believe that $\sigma^{2}_{E} >\sigma^{2}_{W}$. To deal with the variance relationship we specify the following prior distributions:
\begin{eqnarray}
\sigma^{2}_{E} &\sim& Uniform(0,1) \label{eq:prior1}\\
\sigma^{2}_{W}|\sigma^{2}_{E} &\sim& Uniform(0, \sigma^{2}_{E}). \label{eq:prior2}
\end{eqnarray}
Prior distributions (\ref{eq:prior1}) and (\ref{eq:prior2}) ensures the system variance to be less than the observation variance.
Since these are proper prior distributions the resulting posterior distribution will also be proper.\\
\indent In the first stage of calibration, the joint distribution of the observations, states, and unknown parameters is as follows:
\begin{eqnarray*}
\pi({\bf Y}_{1:T}, {\boldsymbol \theta}_{0:T}, \sigma^{2}_{E}, \sigma^{2}_{W}) &=& f_{{\bf Y}}({\bf Y}_{1:T}| {\boldsymbol \theta}_{0:T}, \sigma^{2}_{E}, \sigma^{2}_{W}) \cdot f_{\boldsymbol \theta}({\boldsymbol \theta}_{0:T}| \sigma^{2}_{W}) \cdot \pi({\boldsymbol \Gamma})\\
&=& \prod_{t=1}^{T} f_{{\bf Y}}({\bf Y}_{t}| {\boldsymbol \theta}_{t}, \sigma^{2}_{E}) \cdot \prod_{t=1}^{T} f_{\boldsymbol \theta}({\boldsymbol \theta}_{t}| {\boldsymbol \theta}_{t-1}, \sigma^{2}_{W}) \\ & & \hspace{10pt} \cdot \pi({\boldsymbol \theta}_{0})  \cdot \pi(\sigma^{2}_{E})  \cdot \pi(\sigma^{2}_{W}|\sigma^{2}_{E}).
\end{eqnarray*}
where the likelihood for the observation equation is 
\begin{equation*}
 f_{{\bf Y}}({\bf Y}_{t}|{\boldsymbol \theta}_{t}, \sigma^{2}_{E}) \propto \sigma^{-T}_{E} \mbox{exp}\left\{-\frac{1}{2\sigma^{2}_{E}}\sum_{t=1}^{T}({\bf Y}_{t}-{\bf X}_{t} {\boldsymbol \theta}_{t})^{2}\right\}
\end{equation*}
and the likelihood for the system equation is
\begin{equation*}
 f_{{\boldsymbol \theta}}({\boldsymbol \theta}_{t}|{\boldsymbol \theta}_{t-1}, \sigma^{2}_{W}) \propto \sigma^{-T}_{W}  \mbox{exp}\left\{-\frac{1}{2\sigma^{2}_{W}}\sum_{t=1}^{T}({\boldsymbol \theta}_{t} - {\boldsymbol \theta}_{t-1})^{2}\right\}.
\end{equation*}
Given the joint distribution above, the posterior distribution is
\begin{equation}\label{eq:MOD1post}
\pi({\bf x}_{0t}|{\boldsymbol \theta}_{t}, {\boldsymbol \Gamma}, {\bf Y}_{t})
\end{equation}
where
\begin{equation}\label{eq:calMOD1}
	{\bf x}_{0t} = \frac{\bf{y}^{*}_{0t}}{{\boldsymbol \theta}_{t}}
\end{equation}
and ${\bf y}^{*}_{0t} = {\bf y}_{0t}-\bar{y}_{t}$ (i.e. $\bar{ y}_{t}$ is the cumulative mean of the observations up to time $t$) and ${\boldsymbol \theta}_{t}=\hat{\boldsymbol \beta}_{1t}$.  Samples from the posterior distribution in Equation (\ref{eq:MOD1post}) are drawn by implementing the Sampling Importance Resampling (Albert 2007; Givens and Hoeting 2005) approach.\\  
\indent The development of the estimator in Equation (\ref{eq:calMOD1}) is deterministic in approach. We present a fully Bayesian approach to dynamic calibration that incorporates the uncertainty in estimation. The second dynamic calibration model is derived by Bayes' theorem
\begin{equation*}
\pi({\bf x_{0t}}|{\bf Y}_{t}) \propto \pi({\bf x_{0t}}) f({\bf Y}_{t}| {\bf x_{0t}}),
\end{equation*}
where $\pi({\bf x_{0t}}|{\bf Y}_{t})$ is the posterior distribution for ${\bf x}_{0t}$. The prior belief for the calibration values is denoted as $\pi({\bf x_{0t}})$ with the $f({\bf Y}_{t}| {\bf x_{0t}})$ denoting the likelihood function.\\
\indent The objective of any Bayesian approach is to obtain the posterior distribution from which inferences can be made. Here the desired posterior is
\begin{equation}\label{eq:cali_post}
\pi({\bf x_{0t}}|{\bf Y}_{t})
\end{equation}
which must be dynamic through time. We determine the posterior distribution (\ref{eq:cali_post}) in a similiar manner as described above in Equations (\ref{eq:MOD1post}) and (\ref{eq:calMOD1}).
 ~In the first stage of the calibration experiment the data is scaled and centered, therefore setting the $y-$intercept equal to zero and the reference measurements centered at zero. Centering of the data is used to reduce the parameter space. The posterior distribution can be thought of as:
\begin{equation}\label{eq:center_post}
\pi({\bf z_{0t}}|{\bf Y}^{*}_{t}) \propto \pi({\bf z_{0t}}) f({\bf Y}^{*}_{t}| {\bf z_{0t}}),
\end{equation}
with ${\bf z_{0t}}$ representing the transformed calibrated value at time $t$ and ${\bf Y}^{*}_{t} = {\bf Y}_{t} - \bar{ Y}_{t}$, where $\bar{ Y}_{t}$ is the cumulative mean of the observations. Given this information $a~priori$ we define the prior distribution
\begin{equation*}\label{eq:center_prior}
\pi({\bf z_{0t}}) = N(0,1).
\end{equation*}
The posterior density in Equation (\ref{eq:center_post}) is defined as
\begin{equation}\label{eq:cali_like}
\pi({\bf z_{0t}}|{\bf Y}^{*}_{t}) \propto \mbox{exp}\left\{-\frac{1}{2}\left[\sigma^{-2}_{Y_t}\sum_{t=1}^{T}({\boldsymbol \xi}_{t}-{\bf z}_{0t})^{2}+{\bf z}^{2}_{0t}\right]\right\}
\end{equation} 
where ${\boldsymbol \xi}_{t} =  \nicefrac{{\bf Y}^{*}_{0t}}{\boldsymbol \theta_{t}}$. Applying Bayes theorem and completing the square, the posterior distribution is 
\begin{equation}\label{eq:center_distr}
\pi({\bf z_{0t}}|{\bf Y}^{*}_{t}) \sim N(\mu_{z_{0t}},\sigma^{2}_{z_{0t}}),
\end{equation}
with 
\begin{eqnarray*}
\mu_{z_{0t}} &=& \frac{{\boldsymbol \xi}_{t}}{1+\sigma^{2}_{Y_t}},\\
\sigma^{2}_{z_{0t}} &=& \frac{1}{1+\sigma^{2}_{Y_t}}
\end{eqnarray*}
and 
\begin{equation*}
\sigma^{2}_{Y_t} = \mbox{tr}({\bf Q}_{t}).
\end{equation*}
where tr(~.~) denotes trace of the one-step forecast variance-covariance matrix. We derive the posterior in Equation (\ref{eq:cali_post}) by drawing from Equation (\ref{eq:center_distr}) and transforming the data back to the original scale as so:
\begin{equation}\label{eq:dlm_calib2}
	{\bf x}_{0t} = \bar{X} + {\bf z}_{0t}\sigma_{X},
\end{equation}
where $\bar{X}$ is the mean of the reference measurements vector and $\sigma_{X}$ is the standard deviation of the reference measurements  vector.\\
\indent The dynamic calibration algorithm is developed for both of the approaches using R (R Development Core Team, 2013) and is conducted as below.
\begin{algorithm}\label{DynCal}
{\bf Algorithm:}~Dynamic~Calibration
\small{ 
	\begin{enumerate}
	\item Generate $M$ proposal samples for $(\sigma^{2}_{E}, \sigma^{2}_{W})$ from $\pi(\sigma^{2}_{E})$ and $\pi(\sigma^{2}_{W}|\sigma^{2}_{E})$;
	\item Calibration data are fit using the DLM framework for each of the $M$ proposal samples $(\sigma^{2(m)}_{E}, \sigma^{2(m)}_{W})$, with the prior moments for $({\boldsymbol \theta_{0}} | D_{0})$ as {$\bf m_{0} = 1_{d}$} and {$\bf C_{0} = 100I_{(d \times d)}$}, where ${\bf 1_{d}}$ is a $d-$dimensional vector of ones.\\	
		\begin{enumerate}[a.]
			\item Data are scaled and shifted such that $\sum^{r}_{i=1} x_{i} =0$,  $\frac{1}{n}\sum^{r}_{i=1} x^{2}_{i} =1$ and $y-$intercept $= 0$, where $y^{*}_{t} = y_{t} - \bar{y}_{t}$ for all $t$ (i.e. $\bar{y}_{t}$ is the cumulative mean up to time $t$);
			\item Estimate ${\boldsymbol \theta}^{(m)}_{t}|\sigma^{2(m)}_{E}, \sigma^{2(m)}_{W}$ for the $m^{th}$ proposal sample is calculated $\mbox{for all} ~ t$;
			\item Estimate $x^{(m)}_{0t}|{\boldsymbol \theta}^{(m)}_{t}, \sigma^{2(m)}_{E}, \sigma^{2(m)}_{W}$ for the $m^{th}$ proposal sample is calculated $\mbox{for all} ~ t$, using either Equation (\ref{eq:calMOD1}) or drawing from Equation (\ref{eq:center_distr});
			\item Calculate log-likelihood density weights, $log[f({\boldsymbol \Gamma}^{(m)})]$, for each $(\sigma^{2(m)}_{E}, \sigma^{2(m)}_{W})$ pair
		\end{enumerate}
		\item Sampling Importance Resampling (SIR) is used to simulate  samples of $x_{0t}| {\boldsymbol \theta}_{t},\sigma^{2}_{E}, \sigma^{2}_{W}$ by accepting a subset of $N = 1,000$ from the proposal density to be distributed according to the posterior density $\pi({\boldsymbol \Gamma}|{\bf Y}_{t})$ with candidate density $\pi({\boldsymbol \Gamma})$.
		\begin{enumerate}[a.]
			\item Calculate the standardized importance weights, $w({{\boldsymbol \Gamma}^{(1)}}), \dots, w({{\boldsymbol \Gamma}^{(M)}})$ , where $w({{\boldsymbol \Gamma^{(m)}}}) = log[f({\boldsymbol \Gamma}^{(m)})] - log[g({\boldsymbol \Gamma}^{(m)})]$ for the $m^{th}$ proposal sample;
			\item  Sample $N$ calibrated time series from the $M$ proposal values with replacement given probabilities $p({{\boldsymbol \Gamma}^{(m)}})$ where 
		\begin{equation*}
			p({{\boldsymbol \Gamma}^{(m)}}) = \frac{e^{w({{\boldsymbol \Gamma^{(m)}}})}}{\sum_{j=1}^{M}e^{w({{\boldsymbol \Gamma^{(j)}}})}}.
		\end{equation*}
		\end{enumerate}
	\item Rescale calibrated time series to original scale by Equation (\ref{eq:dlm_calib2}) and take summary statistics (i.e. medians and credible sets) across each time $t$ .
\end{enumerate}
}
\end{algorithm}
%
%
%
%
%
%
%

\section{Simulation Study}\label{sec:Sim_Study}

\begin{figure}
\centering
{\includegraphics[width=2.0in]{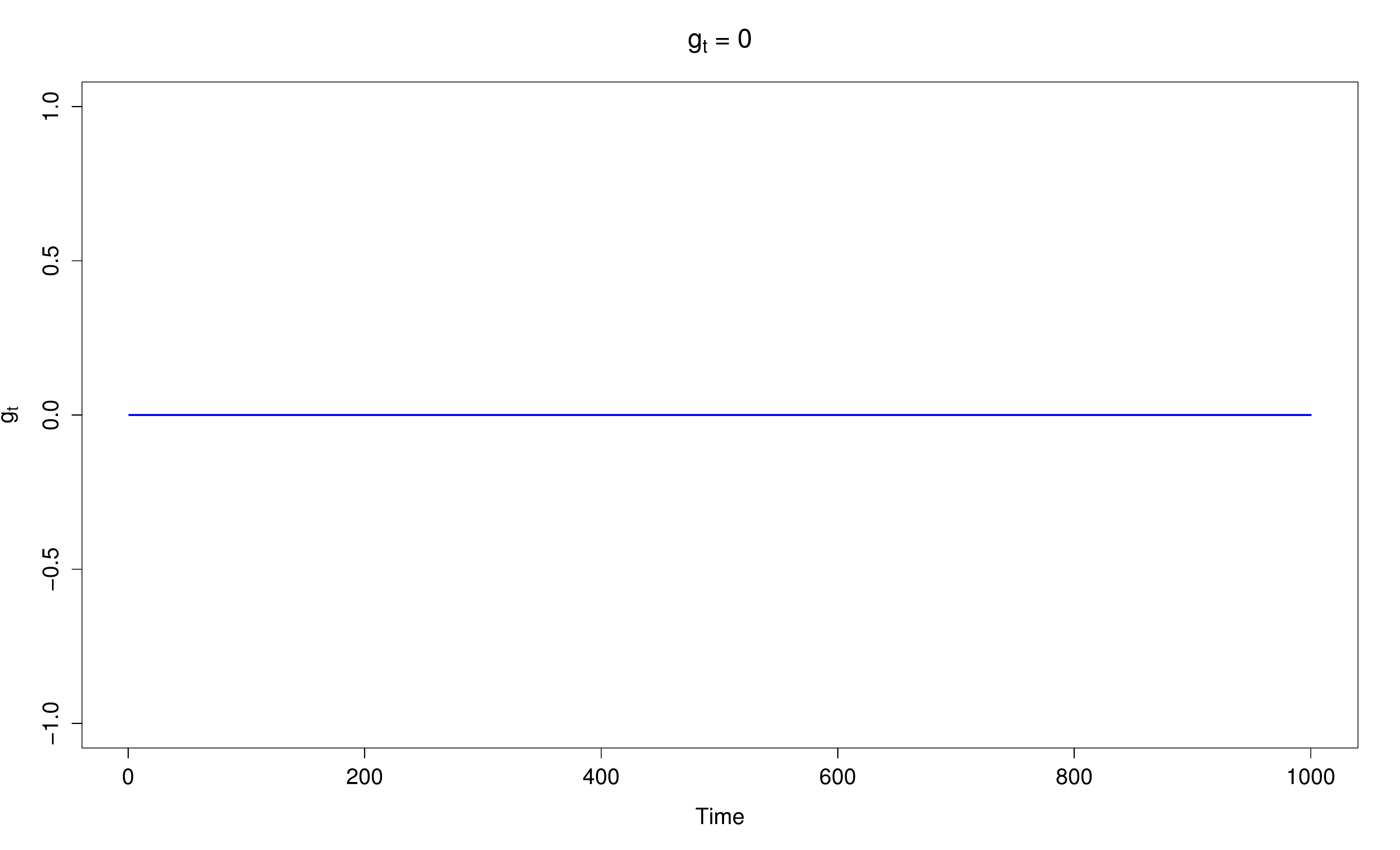}} 
~\quad
{\includegraphics[width=2.0in]{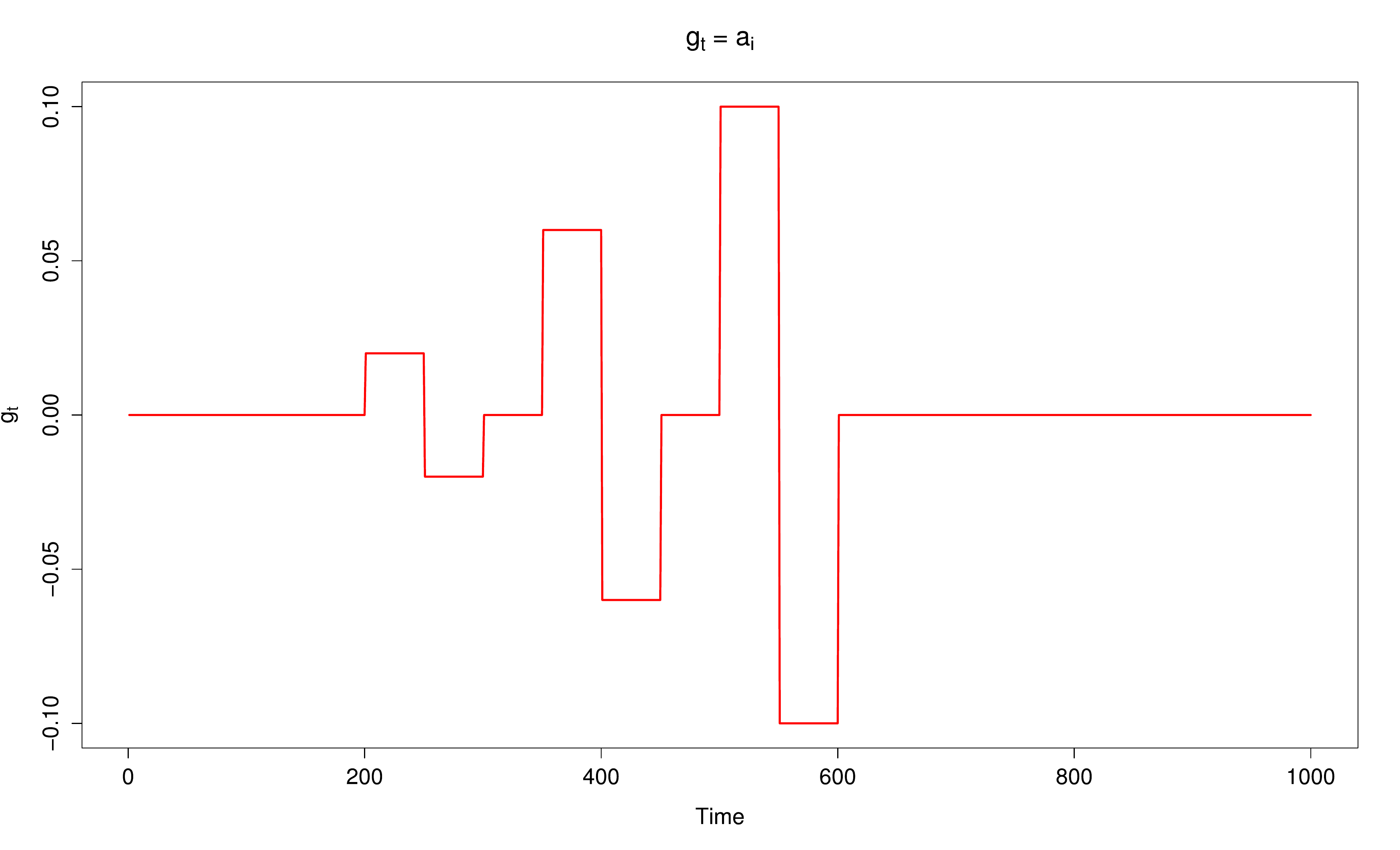}}\\
~
{\includegraphics[width=2.0in]{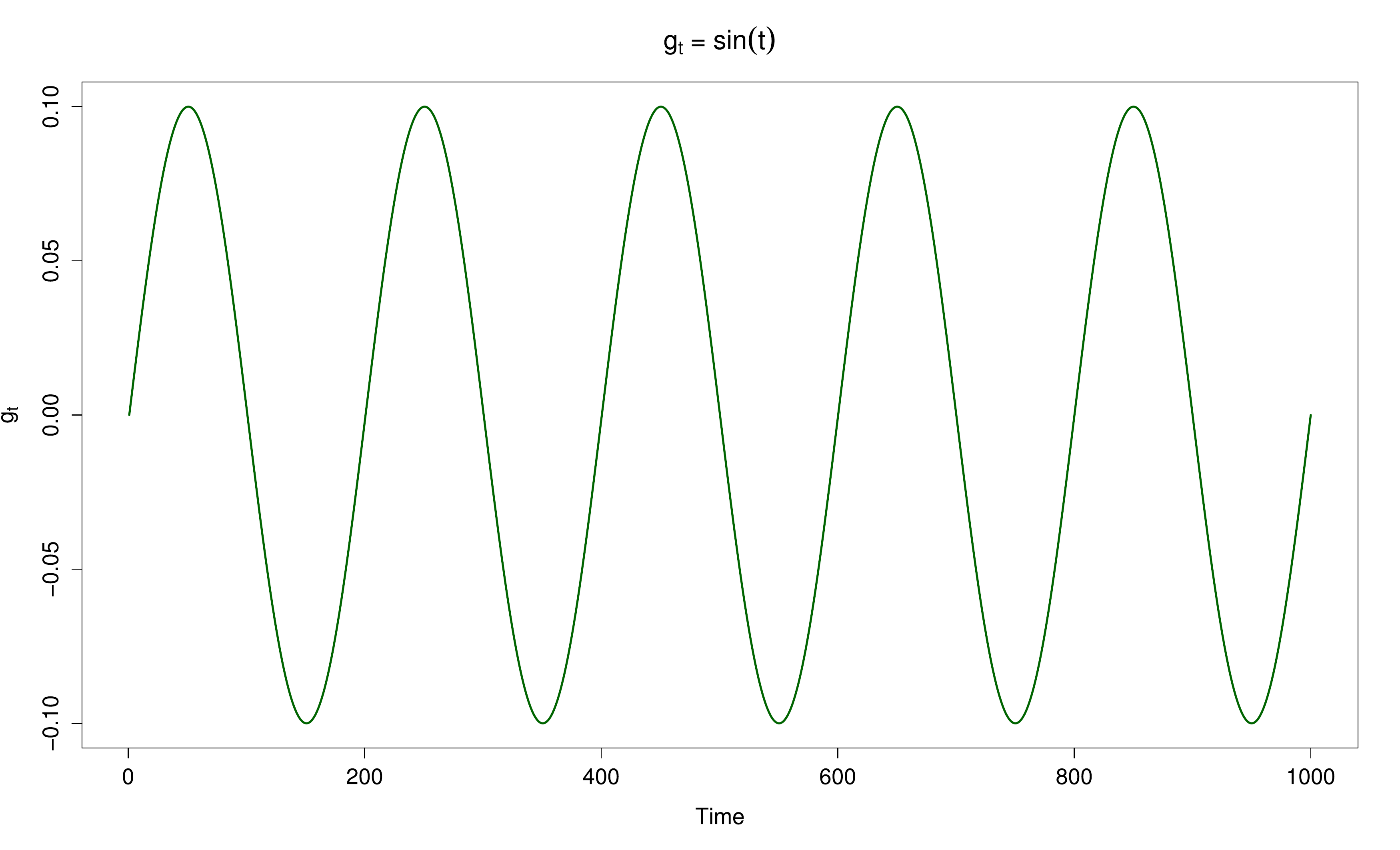}}
     \caption{Three distinct gain fluctuations: (a) $g_{t} = 0$; (b) $g_{t} = a_{i}$ with $\sum_{i=1}^{n}a_{i}=0$; (c) $g_{t} = 0.1\mbox{sin}(0.025t)$}\label{fig:gain_fluct}
\end{figure}

A simulation study, mirroring the microwave radiometer example in Section \ref{sec:application}, considers the performance of the proposed dynamic calibration approaches to the static approaches discussed in Section \ref{sec:Introduction}. For notation, the calibration methods are labelled as follows:
\begin{enumerate}
	\item $M_{D1}$ is the first deterministic dynamic calibration model given in Equation (\ref{eq:calMOD1});
	\item $M_{D2}$ is the Bayesian dynamic calibration model given by Equation (\ref{eq:center_distr});
	\item $M_{F1}$ is the $``classical"$ approach of Eisenhart (1939) defined in Equation (\ref{eq:ClassEq});
	\item $M_{F2}$ is the $``inverse"$ approach of Krutchkoff (1967) defined in Equation (\ref{eq:InvEq});
	\item $M_{B1}$ is Hoadley (1970) Bayesian approach as defined in Equation (\ref{eq:HoadEq});
	\item $M_{B2}$ is Hunter \& Lamboy (1981) Bayesian approach as defined in Equation (\ref{eq:HuntEq}).
\end{enumerate}
\indent Note that static methods $M_{F1}$, $M_{F2}$, $M_{B1}$, and $M_{B2}$ require that model fitting and the calibration take place after all the data has been collected.  This is in contrast to the dynamic methods that both fit the model and generate calibrated values each point through time and hence provide a near real time calibration.  In order to assess the performance of the calibration methods 100 datasets were randomly generated according to 
\begin{equation}\label{eq:sim_model}
{\bf Y}_{t} = {\bf X}{\boldsymbol \theta}_{t} + {\boldsymbol \epsilon}_{t},
\end{equation}
where ${\bf X}$ is a known fixed design matrix of reference values. The number of references measurements used in the study was two and five. The reference values at the first stage  of the simulation study were equally spaced, covering the interval $[20,100]$. For the two reference case, the fixed design matrix is 
\[
{\bf X} = 
	\left[ \begin{array}{cc}
	1 & 20   \\
	1 & 100 \end{array} \right]
\]
and for the five reference case the design matrix is 
\[
{\bf X} = 
	\left[ \begin{array}{cc}
	1 & 20   \\
	1 & 40   \\
	1 & 60   \\
	1 & 80   \\
	1 & 100 \end{array} \right].
\]
The vector of regression parameters, ${\boldsymbol \theta}_{t}$, are randomly drawn from a multivariate normal distribution with mean vector $[ 12.7434 ~0.02655]^{'}$ and variance-covariance matrix, ${\boldsymbol \Sigma} = \sigma^{2}_{W}\left[{\bf X}^{'} {\bf X}\right]^{-1}$ for $t = 1, \dots, T$, where $T=1000$. For each $t$, the random multivariate error vector is
\begin{equation*}
 {\boldsymbol \epsilon}_{t} \sim N_{r}[{\bf 0}, \sigma^{2}_{E}{\bf I}]
\end{equation*}
where the errors are mutually independent. The relationship of the values for $\sigma^{2}_{E}$ and $\sigma^{2}_{W}$ will be explained later.\\
\indent The dynamic and static calibration methods are evaluated for three distinct system fluctuations, $g_{t}$, on the regression slope calculated in the first stage of calibration. The value $g_{t}$ is added to $\beta_{1t}$, therefore making Equation (\ref{eq:sim_model})
\begin{equation*}\label{eq:full_sim_model}
y_{jt} = \beta_{0t} +  (\beta_{1t}+g_{t})x_{j} + \epsilon_{t},\quad t=1,\dots, T\\
\end{equation*}
for the $j^{th}$ calibration references. The three scenarios for the fluctuations $g_{t}$ are as follows: 
\begin{enumerate}
	\item a constant zero ($g_{t}= 0$) for all $t$, representing a stable system; 
	\item a stable system with abrupt shifts ($g_{t}= a_{i})$ in system, with $\sum_{i=1}^{n}a_{i}=0$; and
	\item a constant sinusoidal fluctuation ($g_{t}= 0.1\mbox{sin}(0.025t)$) for all $t$.
\end{enumerate}
Figure \ref{fig:gain_fluct}) explains the relationship of $g_{t}$ across time.\\
 \indent ~The magnitude and relationship of the variance pair $(\sigma^{2}_{E}, \sigma^{2}_{W})$ influence the DLM and hence to study this influence we set the variances to reflect various {\it signal-to-noise ratios}. The true values for $\sigma^{2}_{E}$ and $\sigma^{2}_{W}$ used in the simulation study are $(0.0001, 0.001, 0.01)$ and $(0.00001, 0.00005)$, respectively.
%
%
Petris {\it et al.} (2009) define the signal-to-noise ratio $r$ as follows:
\begin{equation*}
r = \frac{\mbox{Observation Variance}}{\mbox{System Variance}} = \frac{\sigma^{2}_{E}}{\sigma^{2}_{W}}.
\end{equation*}
%
%
The signal-to-noise ratios $r$ in the simulation study were examined in two sets. First, $r$ is set equal to 10, 100, and 1000. Next, the ratio $r$ was set to equal 2, 20, and  200. The variety of $r$ values allow us to examine the methods under different levels of noise. Each simulation is repeated 100 times for both the 2- and 5-point calibration models, thus providing us with 36 possible models for examination from the settings of $r$.\\
\indent After the data was fit with each of the methods we considered the following measures for assessing the performance of the dynamic methods compared to the familiar static approaches: (1) average mean square error; (2) average coverage probability; and (3) average interval width. For each of the simulated data sets, the mean squared error ($MSE$) is calculated as
\begin{equation}
MSE = \frac{1}{T}\sum_{t=1}^{T}(\hat{x}_{0t}-x_{0t})^{2}\nonumber.
\end{equation}
The $MSE$ are averaged across the 100 simulated data  sets thus deriving an average mean squared error ($AvMSE$) as
\begin{equation}
AvMSE = \frac{1}{100}\sum_{j=1}^{100}MSE_{j}\nonumber.
\end{equation}
\indent The coverage probability based on the $95\%$ coverage interval is estimated for all of the calibration methods. The coverage interval for the dynamic and static Bayesian approaches is the $95\%$ credible interval and the $95\%$ confidence interval is used for the frequentist methods. Note, that for credible intervals $x_{0t}^{L}$ is the 0.025 posterior quantile for  $x_{0t}$, and  $x_{0t}^{U}$ is the 0.975 posterior quantile for $x_{0t}$,  where $x_{0t}$ is the true value of the calibration target from the second stage of experimentation, then  $(x_{0t}^{L}, x_{0t}^{U})$ is a $95\%$ credible interval. The coveraged probability ($CP$) is calculated as such
\begin{equation}
 CP = \frac{1}{T}\sum_{t=1}^{T}\psi_{t}\nonumber
\end{equation}
where
\[ \psi_{t} = P[x_{0t}^{L} < x_{0t} < x_{0t}^{U}] = \left\{ \begin{array}{ll}
         0 & \mbox{if $x_{0t} \not\in (x_{0t}^{L}, x_{0t}^{U})$};\\
	 &\\
         1 & \mbox{if $x_{0t} \in (x_{0t}^{L}, x_{0t}^{U})$}.\end{array} \right. \]
The average coverage probability ($AvCP$) is calculated by averaging across the number of replications in the simulation study, where
\begin{equation}
 AvCP = \frac{1}{100}\sum_{j=1}^{100}CP_{j}.\nonumber
\end{equation}
\indent Another quantity of interest to compare the average interval widths ($AvIW$) for the methods, where the average interval widths ($IW$) across the simulated time series is calculated as follows:
\begin{equation*}
IW = \frac{1}{T}\sum_{t=1}^{T}(x_{0t}^{U} - x_{0t}^{L})
\end{equation*}
with the average interval width across the simulation study given as
\begin{equation*}
AvIW = \frac{1}{100}\sum_{j=1}^{100}IW_{j},
\end{equation*}
where $IW_{j}$ is the average interval width for the $j^{th}$ simulation replicate. The performance of the dynamic calibration approaches will be assess using the average coverage probability ($AvCP$), average interval width ($AvIW$) and average mean square ($AvMSE$).\\
\indent We consider the performance of the methods under two conditions: interpolation and extrapolation. Interpolation case is of interest to understand how the method performed when the calibrated time series is within the range of the reference values, [20,~100]. Extrapolation case also conducted to examine the methods when $x_{0t}$ falls outside of the range of the calibration references, where $x_{0t} > 100$. While it not preferable to do extrapolation in the regression case, it is often done in practice in microwave radiometry.\\
\indent All simulations were carried out on the Compile server running $R~ 3.0.2$ (R Development Core Team, 2013) at Virginia Commonwealth University. The Compile server has a Linux OS with 16 CPU cores and 32 GB Ram. Each iteration in the study took approximately 15 minutes with a total of 25.63 hours.

\subsection{Interpolation case}

In the following tables, the simulation results for the dynamic and static calibration methods are provided. The results of simulation studies provide insight into the properties of the calibration approaches. The results in Tables \ref{tab:constant_inter1} and \ref{tab:constant_inter2} indicate that all of the estimators do a good job at approximating the true values of $x_{0t}$ when the gain flucuation $g_{t}$ is set to 0. Even in this case we see as the signal-to-noise ratio $r$ increases so does the $AvMSE$ values. All of the methods have an average coverage probability $AvCP$ of 1 or close. The high coverage rate is of no surprise for a stable system. There does not appear to be an advantage by including more reference measurements (i.e 2- or 5-points) in the model when the system is stable in time. The clear difference is the $AvIW$ values for the dynamic methods compared to the static methods. In Tables \ref{tab:constant_inter1} and \ref{tab:constant_inter2} when $r = 10$ and $r = 2$, the interval for the dynamic methods is wider than those of the four static methods but as $r$ increases the interval width of the dynamic methods remain nearly unchanged as the interval widths for the static methods are 4 to 5 times wider.\\
\indent The simulation results for the stepped gain fluctuations are provided in Tables \ref{tab:stepped_inter1} and \ref{tab:stepped_inter2}. Clearly the presence of the stepped $g_{t}$ has an effect on the fit of the models. The results in Tables \ref{tab:stepped_inter1} show that in nearly all cases, the two dynamic methods $M_{D1}$ and $M_{D2}$ have $AvMSE$ values smaller than the two static Bayesian approaches. The $AvMSE$ values for the dynamic methods are reasonably lower for $r=200$. When $r=1000$, notice the dynamic models $M_{D1}$ and $M_{D2}$ have smaller average mean square errors smaller than the static method $M_{F2}$. The average coverage probability $AvCP$ is comparable for all of the methods and number of references. The dynamic methods consistently have shorter interval widths. The widths of the 95\% credible intervals for $M_{D1}$ and $M_{D2}$ is not affected by the increases in $r$.\\
{ \tiny
\begin{table}[!htb]
\centering
\addtolength{\tabcolsep}{-3pt}
\captionof{table}[Comparison of calibration approaches when interpolating to estimate $x_{0t}$ without gain fluctuations]{
Comparison of calibration approaches when interpolating to estimate $x_{0t}$ without gain fluctuations based on 100 data sets. AvMSE is the average mean squared error, AvCP is the average coverage probability, and AvIW is the average 95\% interval width. The signal-to-noise ratio is denoted as $r$.
}
\begin{tabular}{ccrcr|rcr|rcr}
\multicolumn{11}{c}{Constant $g_{t}=0$}\\
  \toprule
 & &         &   $r=10$   &          &             &     $r=100$  &          &             &   $r=1000$     &        \\ 
  \hline
Ref. &  Model & AvMSE & AvCP & AvIW & AvMSE & AvCP & AvIW & AvMSE & AvCP & AvIW \\ 
  \hline
2 & $M_{D1}$ & 0.0008 & 0.995 & 2.519 & 0.0035 & 0.983 & 2.523 & 0.0307 & 0.939 & 2.517 \\ 
   & $M_{D2}$ & 0.0012 & 1.000 & 3.782 & 0.0038 & 1.000 & 3.782 & 0.0308 & 1.000 & 3.782 \\ 
   & $M_{F1}$ & 0.0001 & 1.000 & 1.224 & 0.0012 & 1.000 & 3.868 & 0.0123 & 1.000 & 12.229 \\ 
   & $M_{F2}$ & 0.0001 & 1.000 & 1.223 & 0.0016 & 1.000 & 3.863 & 0.0335 & 1.000 & 12.168 \\ 
   & $M_{B1}$ & 0.0002 & 0.997 & 1.182 & 0.0022 & 1.000 & 3.866 & 0.0386 & 1.000 & 12.177 \\ 
   & $M_{B2}$ & 0.0014 & 1.000 & 1.458 & 0.0139 & 1.000 & 4.606 & 0.1391 & 1.000 & 14.565 \\ 
  \hline
5& $M_{D1}$ & 0.0008 & 0.995 & 2.496 & 0.0035 & 0.983 & 2.509 & 0.0307 & 0.941 & 2.514 \\ 
  & $M_{D2}$ & 0.0013 & 1.000 & 3.983 & 0.0039 & 1.000 & 3.983 & 0.0307 & 1.000 & 3.983 \\ 
  & $M_{F1}$ & 0.0001 & 1.000 & 1.223 & 0.0012 & 1.000 & 3.865 & 0.0123 & 1.000 & 12.220 \\ 
  & $M_{F2}$ & 0.0001 & 1.000 & 1.222 & 0.0022 & 1.000 & 3.860 & 0.0813 & 1.000 & 12.113 \\ 
  & $M_{B1}$ & 0.0002 & 1.000 & 1.223 & 0.0023 & 1.000 & 3.861 & 0.0792 & 1.000 & 12.116 \\ 
  & $M_{B2}$ & 0.0014 & 1.000 & 1.457 & 0.0139 & 1.000 & 4.604 & 0.1069 & 1.000 & 10.748 \\ 
   \bottomrule
\end{tabular}
\label{tab:constant_inter1}
\end{table}
\begin{table}[!htb]
\centering
\addtolength{\tabcolsep}{-3pt}
\captionof{table}[Comparison of calibration approaches when interpolating to estimate $x_{0t}$ without gain fluctuations]{
Comparison of calibration approaches when interpolating to estimate $x_{0t}$ without gain fluctuations based on 100 data sets. AvMSE is the average mean squared error, AvCP is the average coverage probability, and AvIW is the average 95\% interval width. The signal-to-noise ratio is denoted as $r$.
}
\begin{tabular}{ccrcr|rcr|rcr}
\multicolumn{11}{c}{Constant $g_{t}=0$}\\
  \toprule
& &         &   $r=2$   &          &             &     $r=20$  &          &             &   $r=200$     &        \\ 
  \hline
Ref. &Model & AvMSE & AvCP & AvIW & AvMSE & AvCP & AvIW & AvMSE & AvCP & AvIW \\ 
  \hline
2&$M_{D1}$ & 0.0012 & 0.992 & 2.519 & 0.0041 & 0.981 & 2.520 & 0.0323 & 0.939 & 2.528 \\ 
&$M_{D2}$  & 0.0015 & 1.000 & 3.782 & 0.0044 & 1.000 & 3.782 & 0.0325 & 1.000 & 3.782 \\ 
&$M_{F1}$ & 0.0001 & 1.000 & 1.230 & 0.0010 & 1.000 & 3.871 & 0.0114 & 1.000 & 12.231 \\ 
&$M_{F2}$  & 0.0001 & 1.000 & 1.229 & 0.0012 & 1.000 & 3.866 & 0.0314 & 1.000 & 12.170 \\ 
&$M_{B1}$ & 0.0001 & 1.000 & 1.230 & 0.0019 & 1.000 & 3.869 & 0.0371 & 1.000 & 12.179 \\ 
&$M_{B2}$ & 0.0190 & 1.000 & 1.155 & 0.0243 & 1.000 & 3.767 & 0.1381 & 1.000 & 14.567 \\
  \hline
5&$M_{D1}$ & 0.0011 & 0.992 & 2.508 & 0.0041 & 0.981 & 2.510 & 0.032 & 0.939 & 2.514 \\ 
&$M_{D2}$ & 0.0017 & 1.000 & 3.983 & 0.0045 & 1.000 & 3.983 & 0.032 & 1.000 & 3.983 \\ 
&$M_{F1}$ & 0.0001 & 1.000 & 1.228 & 0.0010 & 1.000 & 3.868 & 0.011 & 1.000 & 12.222 \\ 
&$M_{F2}$ & 0.0001 & 1.000 & 1.227 & 0.0019 & 1.000 & 3.863 & 0.081 & 1.000 & 12.114 \\ 
&$M_{B1}$ & 0.0001 & 1.000 & 1.227 & 0.0021 & 1.000 & 3.864 & 0.082 & 1.000 & 12.118 \\ 
&$M_{B2}$ & 0.0013 & 1.000 & 1.462 & 0.0137 & 1.000 & 4.607 & 0.138 & 1.000 & 14.560 \\ 
   \bottomrule
\end{tabular}
\label{tab:constant_inter2}
\end{table}
\begin{table}[!htb]
\centering
\addtolength{\tabcolsep}{-3pt}
\captionof{table}[Comparison of calibration approaches when interpolating to estimate $x_{0t}$ with stepped gain fluctuations]{
Comparison of calibration approaches when interpolating to estimate $x_{0t}$ with stepped gain fluctuations based on 100 data sets. AvMSE is the average mean squared error, AvCP is the average coverage probability, and AvIW is the average 95\% interval width. The signal-to-noise ratio is denoted as $r$.
}
\begin{tabular}{ccrcr|rcr|rcr}
\multicolumn{11}{c}{Stepped $g_{t} = a_{i}$}\\
  \toprule
& &         &   $r=10$   &          &             &     $r=100$  &          &             &   $r=1000$     &        \\ 
  \hline
 Ref. & Model & AvMSE & AvCP & AvIW & AvMSE & AvCP & AvIW & AvMSE & AvCP & AvIW \\ 
  \hline
2&$M_{D1}$ & 0.0191 & 0.961 & 2.509 & 0.0198 & 0.953 & 2.506 & 0.0406 & 0.926 & 2.543 \\ 
&$M_{D2}$ & 0.0196 & 1.000 & 3.782 & 0.0201 & 1.000 & 3.782 & 0.0408 & 1.000 & 3.783 \\ 
&$M_{F1}$ & 0.0001 & 1.000 & 9.094 & 0.0004 & 1.000 & 9.813 & 0.0094 & 1.000 & 15.209 \\ 
&$M_{F2}$ & 0.0046 & 1.000 & 9.065 & 0.0073 & 1.000 & 9.779 & 0.0528 & 1.000 & 15.098 \\ 
&$M_{B1}$ & 0.0859 & 1.000 & 9.072 & 0.0838 & 1.000 & 9.786 & 0.1866 & 1.000 & 15.109 \\ 
&$M_{B2}$ & 0.1399 & 1.000 & 10.830 & 0.0823 & 1.000 & 11.687 & 0.1836 & 1.000 & 18.115 \\ 
  \hline
5&$M_{D1}$ & 0.0191 & 0.961 & 2.510 & 0.0197 & 0.954 & 2.511 & 0.0405 & 0.924 & 2.516 \\ 
&$M_{D2}$ & 0.0196 & 1.000 & 3.983 & 0.0201 & 1.000 & 3.983 & 0.0405 & 1.000 & 3.983 \\ 
&$M_{F1}$ & 0.0001 & 1.000 & 9.087 & 0.0004 & 1.000 & 9.806 & 0.0094 & 1.000 & 15.199 \\ 
&$M_{F2}$ & 0.0184 & 1.000 & 9.041 & 0.0267 & 1.000 & 9.749 & 0.1620 & 1.000 & 14.995 \\ 
&$M_{B1}$ & 0.0199 & 1.000 & 9.044 & 0.0267 & 1.000 & 9.752 & 0.1559 & 1.000 & 14.999 \\ 
&$M_{B2}$ & 0.0706 & 1.000 & 10.826 & 0.0618 & 1.000 & 8.625 & 0.1742 & 1.000 & 15.091 \\ 
   \bottomrule
\end{tabular}
\label{tab:stepped_inter1}
\end{table}
\begin{table}[!htb]
\centering
\addtolength{\tabcolsep}{-3pt}
\captionof{table}[Comparison of calibration approaches when interpolating to estimate $x_{0t}$ with stepped gain fluctuations]{
Comparison of calibration approaches when interpolating to estimate $x_{0t}$ with stepped gain fluctuations based on 100 data sets. AvMSE is the average mean squared error, AvCP is the average coverage probability, and AvIW is the average 95\% interval width. The signal-to-noise ratio is denoted as $r$.
}
\begin{tabular}{ccrcr|rcr|rcr}
\multicolumn{11}{c}{Stepped $g_{t} = a_{i}$}\\
  \toprule
 & &         &   $r=2$   &          &             &     $r=20$  &          &             &   $r=200$     &        \\ 
  \hline
Ref. & Model & AvMSE & AvCP & AvIW & AvMSE & AvCP & AvIW & AvMSE & AvCP & AvIW \\ 
  \hline
2&$M_{D1}$ & 0.0209 & 0.957 & 2.520 & 0.0219 & 0.950 & 2.522 & 0.0436 & 0.921 & 2.526 \\ 
&$M_{D2}$ & 0.0214 & 1.000 & 3.782 & 0.0222 & 1.000 & 3.782 & 0.0438 & 1.000 & 3.782 \\ 
&$M_{F1}$ & 0.0001 & 1.000 & 9.103 & 0.0003 & 1.000 & 9.822 & 0.0086 & 1.000 & 15.216 \\ 
&$M_{F2}$ & 0.0047 & 1.000 & 9.075 & 0.0073 & 1.000 & 9.788 & 0.0511 & 1.000 & 15.105 \\ 
&$M_{B1}$ & 0.0084 & 1.000 & 9.081 & 0.0115 & 1.000 & 9.795 & 0.0601 & 1.000 & 15.116 \\ 
&$M_{B2}$ & 0.0709 & 1.000 & 10.842 & 0.0826 & 1.000 & 11.698 & 0.2054 & 1.000 & 18.122 \\ 
  \hline
5&$M_{D1}$ & 0.0209 & 0.957 & 2.509 & 0.0218 & 0.949 & 2.511 & 0.0436 & 0.920 & 2.516 \\ 
&$M_{D2}$ & 0.0214 & 1.000 & 3.983 & 0.0221 & 1.000 & 3.983 & 0.0435 & 1.000 & 3.983 \\ 
&$M_{F1}$ & 0.0001 & 1.000 & 9.096 & 0.0003 & 1.000 & 9.815 & 0.0086 & 1.000 & 15.205 \\ 
&$M_{F2}$ & 0.0185 & 1.000 & 9.050 & 0.0267 & 1.000 & 9.758 & 0.1616 & 1.000 & 15.002 \\ 
&$M_{B1}$ & 0.0199 & 1.000 & 9.053 & 0.0281 & 1.000 & 9.761 & 0.1641 & 1.000 & 15.006 \\ 
&$M_{B2}$ & 0.0708 & 1.000 & 10.836 & 0.0825 & 1.000 & 11.693 & 0.2052 & 1.000 & 18.114 \\ 
   \bottomrule
\end{tabular}
\label{tab:stepped_inter2}
\end{table}
\begin{table}[!htb]
\centering
\addtolength{\tabcolsep}{-3pt}
\captionof{table}[Comparison of calibration approaches when interpolating to estimate $x_{0t}$ with sinusoidal gain fluctuations]{
Comparison of calibration approaches when interpolating to estimate $x_{0t}$ with sinusoidal gain fluctuations based on 100 data sets. AvMSE is the average mean squared error, AvCP is the average coverage probability, and AvIW is the average 95\% interval width. The signal-to-noise ratio is denoted as $r$.
}
\begin{tabular}{ccrcr|rcr|rcr}
\multicolumn{11}{c}{Sinusoidal $g_{t} = 0.1\mbox{sin}(0.025t)$}\\
  \toprule
 & &         &   $r=10$   &          &             &     $r=100$  &          &             &   $r=1000$     &        \\ 
  \hline
Ref. & Model & AvMSE & AvCP & AvIW & AvMSE & AvCP & AvIW & AvMSE & AvCP & AvIW \\ 
  \hline
2 & $M_{D1}$ & 4.4088 & 0.628 & 2.657 & 4.4794 & 0.629 & 2.648 & 4.7214 & 0.638 & 2.681 \\ 
  & $M_{D2}$ &  4.4002 & 0.829 & 3.783 & 4.4708 & 0.825 & 3.783 & 4.7123 & 0.810 & 3.783 \\ 
  & $M_{F1}$ & 0.0001 & 1.000 & 21.980 & 0.0012 & 1.000 & 22.307 & 0.0123 & 1.000 & 25.206 \\ 
  & $M_{F2}$ & 0.1541 & 1.000 & 21.665 & 0.1670 & 1.000 & 21.978 & 0.2943 & 1.000 & 24.738 \\ 
&$M_{B1}$ & 0.1689 & 0.975 & 20.933 & 0.1868 & 1.000 & 21.994 & 0.3174 & 1.000 & 24.757 \\ 
&$M_{B2}$ & 0.4127 & 1.000 & 26.178 & 0.4258 & 1.000 & 26.567 & 0.5531 & 1.000 & 30.020 \\  
  \hline
5&$M_{D1}$ & 4.4087 & 0.628 & 2.646 & 4.4793 & 0.630 & 2.648 & 4.7214 & 0.635 & 2.653 \\ 
&$M_{D2}$ & 4.3906 & 0.845 & 3.984 & 4.4609 & 0.839 & 3.984 & 4.7023 & 0.824 & 3.984 \\ 
&$M_{F1}$ & 0.0001 & 1.000 & 21.964 & 0.0012 & 1.000 & 22.291 & 0.0123 & 1.000 & 25.188 \\ 
&$M_{F2}$ & 0.5810 & 1.000 & 21.371 & 0.6218 & 1.000 & 21.671 & 1.0152 & 1.000 & 24.306 \\ 
&$M_{B1}$ & 0.5956 & 1.000 & 21.377 & 0.5909 & 1.000 & 21.678 & 0.9628 & 1.000 & 24.314 \\ 
&$M_{B2}$ & 0.4123 & 1.000 & 26.166 & 0.3087 & 1.000 & 18.973 & 0.4658 & 1.000 & 25.009 \\ 
   \bottomrule
\end{tabular}
\label{tab:sine_inter1}
\end{table}
\begin{table}[!htb]
\centering
\addtolength{\tabcolsep}{-3pt}
\captionof{table}[Comparison of calibration approaches when interpolating to estimate $x_{0t}$ with sinusoidal gain fluctuations]{
Comparison of calibration approaches when interpolating to estimate $x_{0t}$ with sinusoidal gain fluctuations based on 100 data sets. AvMSE is the average mean squared error, AvCP is the average coverage probability, and AvIW is the average 95\% interval width. The signal-to-noise ratio is denoted as $r$.
}
\begin{tabular}{ccrcr|rcr|rcr}
\multicolumn{11}{c}{Sinusoidal $g_{t} = 0.1\mbox{sin}(0.025t)$}\\
  \toprule
 & &         &   $r=2$   &          &             &     $r=20$  &          &             &   $r=200$     &        \\ 
  \hline
Ref. & Model & AvMSE & AvCP & AvIW & AvMSE & AvCP & AvIW & AvMSE & AvCP & AvIW \\ 
  \hline
2&$M_{D1}$ & 4.4504 & 0.625 & 2.658 & 4.5213 & 0.628 & 2.660 & 4.7643 & 0.6331 & 2.665 \\ 
&$M_{D2}$ & 4.4419 & 0.828 & 3.783 & 4.5127 & 0.823 & 3.783 & 4.7553 & 0.8083 & 3.783 \\ 
&$M_{F1}$ & 0.0001 & 1.000 & 21.968 & 0.0010 & 1.000 & 22.295 & 0.0114 & 1.000 & 25.196 \\ 
&$M_{F2}$ & 0.1538 & 1.000 & 21.653 & 0.1672 & 1.000 & 21.966 & 0.2896 & 1.000 & 24.729 \\ 
&$M_{B1}$ & 0.1732 & 1.000 & 21.669 & 0.1867 & 1.000 & 21.982 & 0.3127 & 1.000 & 24.748 \\ 
&$M_{B2}$ & 0.4122 & 1.000 & 26.164 & 0.4253 & 1.000 & 26.553 & 0.5518 & 1.000 & 30.008 \\
  \hline
5&$M_{D1}$ & 4.4504 & 0.625 & 2.647 & 4.5213 & 0.627 & 2.648 & 4.7643 & 0.633 & 2.654 \\ 
&$M_{D2}$ & 4.4322 & 0.844 & 3.984 & 4.5029 & 0.838 & 3.984 & 4.7451 & 0.823 & 3.984 \\ 
&$M_{F1}$ & 0.0001 & 1.000 & 21.952 & 0.0010 & 1.000 & 22.278 & 0.0114 & 1.000 & 25.178 \\ 
&$M_{F2}$ & 0.5799 & 1.000 & 21.359 & 0.6206 & 1.000 & 21.660 & 1.0133 & 1.000 & 24.297 \\ 
&$M_{B1}$ & 0.5851 & 1.000 & 21.366 & 0.6256 & 1.000 & 21.667 & 1.0183 & 1.000 & 24.304 \\ 
&$M_{B2}$ & 0.4119 & 1.000 & 26.152 & 0.4247 & 1.000 & 26.541 & 0.5513 & 1.000 & 29.995 \\ 
   \bottomrule
\end{tabular}
\label{tab:sine_inter2}
\end{table}
}
\indent The results provided in Tables \ref{tab:sine_inter1} and \ref{tab:sine_inter2} summarize the performance of the methods when the gain fluctuation is sinusoidal noise. The results for $r$ values of 10, 100, and 1000 are given in Table \ref{tab:sine_inter1} with $r=2,~20$ and 200 given in Table \ref{tab:sine_inter2}. When $g_{t}$ is sinusoidal, the $AvMSE$ values for the dynamic methods are consistently larger than any of the static methods. For all of the chosen $r$ values, the $AvCP$ is considerably lower than the opposing methods. The dynamic methods still have average interval widths extremely shorter than any of the static methods. The $AvIW$ is constant across the signal-to-noise ratios.\\ 
\indent The simulation study shows that methods $M_{D1}$ and $M_{D2}$ do a good job at estimating calibrated values that are interior to the range of reference measurements. Both methods display high coverage probabilities in the presence of drifting parameters. For the three possible gain fluctuations, the interval widths for the dynamic methods were consistently shorter than the static calibration approaches. When fitting data where there is a definite linear relationship the dynamic methods are invariant to the number of reference measurements. When using the proposed methods in this paper, not much will be gained by using more than 2 reference measurements. Overall, when interpolating to estimate $x_{0t}$, the dynamic methods outperform the static Bayesian approaches across the different signal-to-noise ratios. In the following section the performance of the dynamic methods are assessed when the calibrated values fall outside of the range of reference measurements.



\subsection{Extrapolation case}
	
At this point in the paper we examine the calibration approaches when the calibrated values are outside of the reference measurements. The range of the measurement references is from 20 to 100. The true $x_{0t}$ behaved as a random walk bounded between 100 and 110. We assessed the performance of the dynamic methods under three possible gain fluctuation patterns. First, the simulation study is conducted without the presence of additional gain fluctuation (i.e. $g_{t}=0$); second, the gain $g_{t}$ is a stepped pattern influencing the time-varying slope $\beta_{1t}$ over time; lastly, a sinusoidal $g_{t}$ is added to $\beta_{1t}$. Just as the previous results, the methods are assessed by the average mean square error ($AvMSE$), average coverage probability ($AvCP$), and the average interval width ($AvIW$) under different signal-noise-ratios.\\ 
{ \tiny
\begin{table}[!htb]
\centering
\addtolength{\tabcolsep}{-3pt}
\captionof{table}[Comparison of calibration approaches when extrapolating to estimate $x_{0t}$ without gain fluctuations]{
Comparison of calibration approaches when extrapolating to estimate $x_{0t}$ without gain fluctuations based on 100 data sets. AvMSE is the average mean squared error, AvCP is the average coverage probability, and AvIW is the average 95\% interval width. The signal-to-noise ratio is denoted as $r$.
}
\begin{tabular}{ccrcr|rcr|rcr}
\multicolumn{11}{c}{Constant $g_{t} = 0$}\\
  \toprule
 &  &         &   $r=10$   &          &             &     $r=100$  &          &             &   $r=1000$     &        \\  
  \hline
Ref. & Model & AvMSE & AvCP & AvIW & AvMSE & AvCP & AvIW & AvMSE & AvCP & AvIW \\ 
  \hline
2&$M_{D1}$ & 0.0018 & 1.000 & 5.255 & 0.0043 & 1.000 & 5.255 & 0.0309 & 1.000 & 5.255 \\ 
&$M_{D2}$ & 0.0016 & 1.000 & 3.910 & 0.0042 & 1.000 & 3.910 & 0.0311 & 1.000 & 3.910 \\ 
&$M_{F1}$ & 0.0001 & 1.000 & 1.224 & 0.0012 & 1.000 & 3.869 & 0.0123 & 1.000 & 12.234 \\ 
&$M_{F2}$ & 0.0001 & 1.000 & 1.223 & 0.0001 & 1.000 & 3.863 & 0.1019 & 1.000 & 12.168 \\ 
&$M_{B1}$ & 0.0001 & 1.000 & 1.225 & 0.0008 & 1.000 & 3.867 & 0.1115 & 1.000 & 12.181 \\ 
&$M_{B2}$ & 0.0014 & 1.000 & 1.458 & 0.0139 & 1.000 & 4.606 & 0.1391 & 1.000 & 14.565 \\ 
  \hline
5&$M_{D1}$ & 0.0019 & 1.000 & 5.233 & 0.0043 & 1.000 & 5.233 & 0.0309 & 1.000 & 5.233 \\ 
&$M_{D2}$ & 0.0029 & 1.000 & 4.106 & 0.0054 & 1.000 & 4.106 & 0.0323 & 1.000 & 4.106 \\ 
&$M_{F1}$ & 0.0001 & 1.000 & 1.223 & 0.0012 & 1.000 & 3.866 & 0.0123 & 1.000 & 12.224 \\ 
&$M_{F2}$ & 0.0001 & 1.000 & 1.222 & 0.0027 & 1.000 & 3.860 & 0.5502 & 1.000 & 12.113 \\ 
&$M_{B1}$ & 0.0001 & 1.000 & 1.223 & 0.0030 & 1.000 & 3.862 & 0.5566 & 1.000 & 12.120 \\ 
&$M_{B2}$ & 0.0014 & 1.000 & 1.457 & 0.0139 & 1.000 & 4.604 & 0.1389 & 1.000 & 14.558 \\ 
   \bottomrule
\end{tabular}
\label{tab:constant_extrap1}
\end{table}
\begin{table}[!htb]
\centering
\addtolength{\tabcolsep}{-3pt}
\captionof{table}[Comparison of calibration approaches when extrapolating to estimate $x_{0t}$ without gain fluctuations]{
Comparison of calibration approaches when extrapolating to estimate $x_{0t}$ without gain fluctuations based on 100 data sets. AvMSE is the average mean squared error, AvCP is the average coverage probability, and AvIW is the average 95\% interval width. The signal-to-noise ratio is denoted as $r$.
}
\begin{tabular}{ccrcr|rcr|rcr}
\multicolumn{11}{c}{Constant $g_{t} = 0$}\\
  \toprule
 &  &         &   $r=2$   &          &             &     $r=20$  &          &             &   $r=200$     &        \\  
  \hline
Ref. & Model & AvMSE & AvCP & AvIW & AvMSE & AvCP & AvIW & AvMSE & AvCP & AvIW \\ 
  \hline
2&$M_{D1}$ & 0.0031 & 1.000 & 5.253 & 0.0060 & 1.000 & 5.253 & 0.0340 & 1.000 & 5.253 \\ 
&$M_{D2}$ & 0.0034 & 1.000 & 3.910 & 0.0064 & 1.000 & 3.910 & 0.0347 & 1.000 & 3.910 \\ 
&$M_{F1}$ & 0.0001 & 1.000 & 1.230 & 0.0008 & 1.000 & 3.872 & 0.0107 & 1.000 & 12.236 \\ 
&$M_{F2}$ & 0.0001 & 1.000 & 1.229 & 0.0003 & 1.000 & 3.866 & 0.1068 & 1.000 & 12.170 \\ 
&$M_{B1}$ & 0.0001 & 1.000 & 1.230 & 0.0010 & 1.000 & 3.871 & 0.1164 & 1.000 & 12.183 \\ 
&$M_{B2}$ & 0.0013 & 1.000 & 1.465 & 0.0135 & 1.000 & 4.610 & 0.1376 & 1.000 & 14.567 \\ 
  \hline
5&$M_{D1}$ & 0.0032 & 1.000 & 5.231 & 0.0060 & 1.000 & 5.231 & 0.0340 & 1.000 & 5.232 \\ 
&$M_{D2}$ & 0.0053 & 1.000 & 4.106 & 0.0083 & 1.000 & 4.106 & 0.0365 & 1.000 & 4.106 \\ 
&$M_{F1}$ & 0.0001 & 1.000 & 1.228 & 0.0008 & 1.000 & 3.869 & 0.0107 & 1.000 & 12.226 \\ 
&$M_{F2}$ & 0.0001 & 1.000 & 1.227 & 0.0035 & 1.000 & 3.863 & 0.5616 & 1.000 & 12.114 \\ 
&$M_{B1}$ & 0.0001 & 1.000 & 1.228 & 0.0039 & 1.000 & 3.865 & 0.5680 & 1.000 & 12.121 \\ 
&$M_{B2}$ & 0.0013 & 1.000 & 1.462 & 0.0135 & 1.000 & 4.607 & 0.1375 & 1.000 & 14.560 \\ 
   \bottomrule
\end{tabular}
\label{tab:constant_extrap2}
\end{table}
}
\indent The results are provided in Tables \ref{tab:constant_extrap1} and\ref{tab:constant_extrap2} for the statistical calibration methods without gain fluctuations. The performance of the proposed method is stable across the signal-to-noise ratios. A point of interest is the reported $AvIW$ values for methods $M_{D1}$ and $M_{D2}$. We see for $r=10$ and $r=2$ that the $AvIW$ is 3 to 5 times wider than those for the static approaches. When $r=100$ and $r=20$ the interval width for all competing methods are relatively close. The dynamic approaches outperform the static methods in noisy conditions such as $r=1000$ and $r=200$. The interval widths for the dynamic methods are considerably shorter than the those for the static methods. The simulation results reveal that when the data is characteristic of having a large signal-to-noise ratio, the dynamic methods, $M_{D1}$ and $M_{D2}$, will outperform static Bayesian approaches and the inverse approach.\\
{\tiny
\begin{table}[!htb]
\centering
\addtolength{\tabcolsep}{-3pt}
\captionof{table}[Comparison of calibration approaches when extrapolating to estimate $x_{0t}$ with stepped gain fluctuations]{
Comparison of calibration approaches when extrapolating to estimate $x_{0t}$ with stepped gain fluctuations based on 100 data sets. AvMSE is the average mean squared error, AvCP is the average coverage probability, and AvIW is the average 95\% interval width. The signal-to-noise ratio is denoted as $r$.
}
\begin{tabular}{ccrcr|rcr|rcr}
\multicolumn{11}{c}{Stepped $g_{t} = a_{i}$}\\
  \toprule
 &  &         &   $r=10$   &          &             &     $r=100$  &          &             &   $r=1000$     &        \\  
  \hline
Ref. & Model & AvMSE & AvCP & AvIW & AvMSE & AvCP & AvIW & AvMSE & AvCP & AvIW \\ 
  \hline
2&$M_{D1}$ & 0.0206 & 1.000 & 5.247 & 0.0210 & 1.000 & 5.247 & 0.0412 & 1.000 & 5.247 \\ 
&$M_{D2}$ & 0.0225 & 1.000 & 3.910 & 0.0230 & 1.000 & 3.910 & 0.0435 & 1.000 & 3.910 \\ 
&$M_{F1}$ & 0.0001 & 1.000 & 9.097 & 0.0004 & 1.000 & 9.817 & 0.0094 & 1.000 & 15.215 \\ 
&$M_{F2}$ & 0.0581 & 1.000 & 9.065 & 0.0656 & 1.000 & 9.779 & 0.3191 & 1.000 & 15.098 \\ 
&$M_{B1}$ & 0.0634 & 1.000 & 9.075 & 0.0718 & 1.000 & 9.789 & 0.3361 & 1.000 & 15.115 \\ 
&$M_{B2}$ & 0.0707 & 1.000 & 10.830 & 0.0826 & 1.000 & 11.687 & 0.2060 & 1.000 & 18.115 \\ 
  \hline
5&$M_{D1}$ & 0.0209 & 1.000 & 5.226 & 0.0213 & 1.000 & 5.226 & 0.0412 & 1.000 & 5.226 \\ 
&$M_{D2}$ & 0.0268 & 1.000 & 4.106  & 0.0273 & 1.00 & 4.106 & 0.0483 & 1.000 & 4.106 \\ 
&$M_{F1}$ & 0.0001 & 1.000 & 9.090 & 0.0004 & 1.000 & 9.809 & 0.0094 & 1.000 & 15.203 \\ 
&$M_{F2}$ & 0.2274 & 1.000 & 9.041 & 0.2812 & 1.000 & 9.749 & 1.4628 & 1.000 & 14.995 \\ 
&$M_{B1}$ & 0.2307 & 1.000 & 9.047 & 0.2851 & 1.000 & 9.755 & 1.4744 & 1.000 & 15.004 \\ 
&$M_{B2}$ & 0.0706 & 1.000 & 10.826 & 0.0825 & 1.000 & 11.682 & 0.2058 & 1.000 & 18.106 \\ 
   \bottomrule
\end{tabular}
\label{tab:stepped_extrap1}
\end{table}
}
{\tiny
\begin{table}[!htb]
\centering
\addtolength{\tabcolsep}{-3pt}
\captionof{table}[Comparison of calibration approaches when extrapolating to estimate $x_{0t}$ with stepped gain fluctuations]{
Comparison of calibration approaches when extrapolating to estimate $x_{0t}$ with stepped gain fluctuations based on 100 data sets. AvMSE is the average mean squared error, AvCP is the average coverage probability, and AvIW is the average 95\% interval width. The signal-to-noise ratio is denoted as $r$.
}
\begin{tabular}{ccrcr|rcr|rcr}
\multicolumn{11}{c}{Stepped $g_{t} = a_{i}$}\\
  \toprule
 &  &         &   $r=2$   &          &             &     $r=20$  &          &             &   $r=200$     &        \\  
  \hline
Ref. & Model & AvMSE & AvCP & AvIW & AvMSE & AvCP & AvIW & AvMSE & AvCP & AvIW \\ 
  \hline
2&$M_{D1}$ & 0.0242 & 1.000 & 5.245 & 0.0250 & 1.000 & 5.245 & 0.0466 & 1.000 & 5.245 \\ 
&$M_{D2}$ & 0.0266 & 1.000 & 3.910 & 0.0275 & 1.000 & 3.910 & 0.0494 & 1.000 &  3.910 \\ 
&$M_{F1}$ & 0.0001 & 1.000 & 9.106 & 0.0002 & 1.000 & 9.826 & 0.0080 & 1.000 & 15.222 \\ 
&$M_{F2}$ & 0.0620 & 1.000 & 9.075 & 0.0698 & 1.000 & 9.788 & 0.3284 & 1.000 & 15.105 \\ 
&$M_{B1}$ & 0.0674 & 1.000 & 9.085 & 0.0760 & 1.000 & 9.799 & 0.3447 & 1.000 & 15.121 \\ 
&$M_{B2}$ & 0.0710 & 1.000 & 10.842 & 0.0825 & 1.000 & 11.698 & 0.2048 & 1.000 & 18.122 \\ 
  \hline
5&$M_{D1}$ & 0.0245 & 1.000 & 5.224 & 0.0254 &  1.000 & 5.224 & 0.0427 & 1.000 & 5.226 \\ 
&$M_{D2}$ & 0.0315 & 1.000 & 4.106 & 0.0324 & 1.000 & 4.106 & 0.0485 & 1.000 & 4.106 \\ 
&$M_{F1}$ & 0.0001 & 1.000 & 9.099 & 0.0002 & 1.000 & 9.818 & 0.0089 & 1.000 & 14.255 \\ 
&$M_{F2}$ & 0.2354 & 1.000 & 9.050 & 0.2902 & 1.000 & 9.758 & 1.1896 & 1.000 & 14.078 \\ 
&$M_{B1}$ & 0.2388 & 1.000 & 9.056 & 0.2941 & 1.000 & 9.764 & 1.1995 & 1.000 & 14.086 \\ 
&$M_{B2}$ & 0.0709 & 1.000 & 10.836 & 0.0824 & 1.000 & 11.693 & 0.1831 & 1.000 & 16.977 \\ 
   \bottomrule
\end{tabular}
\label{tab:stepped_extrap2}
\end{table}
}
{\tiny
%
\begin{table}[!htb]
\centering
\addtolength{\tabcolsep}{-3pt}
\captionof{table}[Comparison of calibration approaches when extrapolating to estimate $x_{0t}$ with sinusoidal gain fluctuations]{
Comparison of calibration approaches when extrapolating to estimate $x_{0t}$ with sinusoidal gain fluctuations based on 100 data sets. AvMSE is the average mean squared error, AvCP is the average coverage probability, and AvIW is the average 95\% interval width. The signal-to-noise ratio is denoted as $r$.
}
\begin{tabular}{ccrcr|rcr|rcr}
\multicolumn{11}{c}{Sinusoidal $g_{t} = 0.1\mbox{sin}(0.025t)$}\\
  \toprule
 &  &         &   $r=10$   &          &             &     $r=100$  &          &             &   $r=1000$     &        \\  
  \hline
Ref. & Model & AvMSE & AvCP & AvIW & AvMSE & AvCP & AvIW & AvMSE & AvCP & AvIW \\ 
  \hline
2&$M_{D1}$ & 4.4096 & 0.873 & 5.127 & 4.4800 & 0.872 & 5.127 & 4.7214 & 0.866 & 5.127 \\ 
&$M_{D2}$ & 4.4410 & 0.833 & 3.904 & 4.5114 & 0.825 & 3.904 & 4.7530 & 0.813 & 3.904 \\ 
&$M_{F1}$ & 0.0001 & 1.000 & 21.988 & 0.0012 & 1.000 & 22.315 & 0.0123 & 1.000 & 25.216 \\ 
&$M_{F2}$ & 1.8193 & 1.000 & 21.665 & 1.8636 & 1.000 & 21.978 & 2.7760 & 1.000 & 24.739 \\ 
&$M_{B1}$ & 1.8602 & 1.000 & 21.688 & 1.9056 & 1.000 & 22.002 & 2.8312 & 1.000 & 24.766 \\ 
&$M_{B2}$ & 0.4127 & 1.000 & 26.178 & 0.4258 & 1.000 & 26.567 & 0.5531 & 1.000 & 30.020 \\ 
  \hline
5&$M_{D1}$ & 4.4105 & 0.872 & 5.106 & 4.4808 & 0.872 & 5.106 & 4.7222 & 0.866 & 5.107 \\ 
&$M_{D2}$ & 4.4889 & 0.842 & 4.100 & 4.5593 & 0.835 & 4.101 & 4.8007 & 0.822 & 4.100 \\ 
&$M_{F1}$ & 0.0001 & 1.000 & 21.971 & 0.0012 & 1.000 & 22.297 & 0.0123 & 1.000 & 25.195 \\ 
&$M_{F2}$ & 6.9539 & 1.000 & 21.371 & 7.2327 & 1.000 & 21.671 & 11.0337 & 1.000 & 24.306 \\ 
&$M_{B1}$ & 6.9852 & 1.000 & 21.383 & 7.2650 & 1.000 & 21.684 & 11.0772 & 1.000 & 24.320 \\ 
&$M_{B2}$ & 0.4123 & 1.000 & 26.166 & 0.4254 & 1.000 & 26.555 & 0.5526 & 1.000 & 30.007 \\ 
   \bottomrule
\end{tabular}
\label{tab:sine_extrap1}
\end{table}
\begin{table}[!htb]
\centering
\addtolength{\tabcolsep}{-3pt}
\captionof{table}[Comparison of calibration approaches when extrapolating to estimate $x_{0t}$ with sinusoidal gain fluctuations]{
Comparison of calibration approaches when extrapolating to estimate $x_{0t}$ with sinusoidal gain fluctuations based on 100 data sets. AvMSE is the average mean squared error, AvCP is the average coverage probability, and AvIW is the average 95\% interval width. The signal-to-noise ratio is denoted as $r$.
}
\begin{tabular}{ccrcr|rcr|rcr}
\multicolumn{11}{c}{Sinusoidal $g_{t} = 0.1\mbox{sin}(0.025t)$}\\
  \toprule
 &  &         &   $r=2$   &          &             &     $r=20$  &          &             &   $r=200$     &        \\  
  \hline
Ref. & Model & AvMSE & AvCP & AvIW & AvMSE & AvCP & AvIW & AvMSE & AvCP & AvIW \\ 
  \hline
2&$M_{D1}$ & 4.4491 & 0.872 & 5.125 & 4.5199 & 0.871 & 5.125 & 4.7626 & 0.866 & 5.126 \\ 
&$M_{D2}$ & 4.4809 & 0.828 & 3.904 & 4.5518 & 0.821 & 3.904 & 4.7948 & 0.807 & 3.904 \\ 
&$M_{F1}$ & 0.0001 & 1.000 & 21.976 & 0.0008 & 1.000 & 22.303 & 0.0107 & 1.000 & 25.205 \\ 
&$M_{F2}$ & 1.8350 & 1.000 & 21.653 & 1.8796 & 1.000 & 21.966 & 2.7956 & 1.000 & 24.729 \\ 
&$M_{B1}$ & 1.8759 & 1.000 & 21.676 & 1.9216 & 1.000 & 21.990 & 2.8508 & 1.000 & 24.756 \\ 
&$M_{B2}$ & 0.4123 & 1.000 & 26.164 & 0.4250 & 1.000 & 26.553 & 0.5511 & 1.000 & 30.008 \\ 
  \hline
5&$M_{D1}$ & 4.4497 & 0.872 & 5.10 & 4.5205 & 0.871 & 5.105 & 4.7633 & 0.865 & 5.105 \\ 
&$M_{D2}$ & 4.5292 & 0.836 & 4.100 & 4.6000 & 0.832 & 4.100 & 4.842 & 0.814 & 4.100 \\ 
&$M_{F1}$ & 0.0001 & 1.000 & 21.958 & 0.0008 & 1.000 & 22.285 & 0.0107 & 1.000 & 25.185 \\ 
&$M_{F2}$ & 6.9764 & 1.000 & 21.359 & 7.2560 & 1.000 & 21.660 & 11.0629 & 1.000 & 24.297 \\ 
&$M_{B1}$ & 7.0077 & 1.000 & 21.372 & 7.2882 & 1.000 & 21.673 & 11.1064 & 1.000 & 24.311 \\ 
&$M_{B2}$ & 0.4119 & 1.000 & 26.152 & 0.4246 & 1.000 & 26.541 & 0.5507 & 1.000 & 29.995 \\ 
   \bottomrule
\end{tabular}
\label{tab:sine_extrap2}
\end{table}
}
\indent Next, we impose a stepped gain fluctuation $g_{t}$ to the data generated and wanted to evaluate the behavior of the calibration methods. The results for the stepped case are given in Tables \ref{tab:stepped_extrap1} and \ref{tab:stepped_extrap2}. We see by the $AvMSE$ values in both tables that the dynamic methods perform better than most static methods. If the calibrated values by chance drift outside of the reference range the dynamic methods will do a good job at capturing it with certainty while having a narrower credible interval than confidence intervals of the static methods. The dynamic approaches outperform all of the static method in terms of $AvIW$. These results of the simulation study do not change much across the number of references used. Once again, when the relationship is assumed to be linear there is no benefit to adding more references.\\
\indent Lastly, the study is conducted with a sinusoidal gain fluctuation while extrapolating to estimate $x_{0t}$. The results for the sinusoidal case are given in Tables \ref{tab:sine_extrap1} and \ref{tab:sine_extrap2}. The dynamic methods $M_{D1}$ and $M_{D2}$ exhibit the same behavior as before in Tables  \ref{tab:sine_inter1} and \ref{tab:sine_inter2} with $AvMSE$ values ranging for 4.4 to 4.8. Even though the average mean square errors are larger than those of the static methods when using a 2-reference model, the two dynamic methods outperform the static methods $M_{F2}$ and $M_{B1}$ which are based on the inverse approach. The dynamic models have average coverage probabilities smaller than the static model across all of the signal-to-noise ratios.  We can not fail to point out that once again the $AvIW$ are 4 to 6 times shorter than the average widths for the static models.

%

	\section{Application to Microwave Radiometer}\label{sec:application}

\begin{figure}[!h]
\centering
\includegraphics[width = 9cm, height = 4cm]{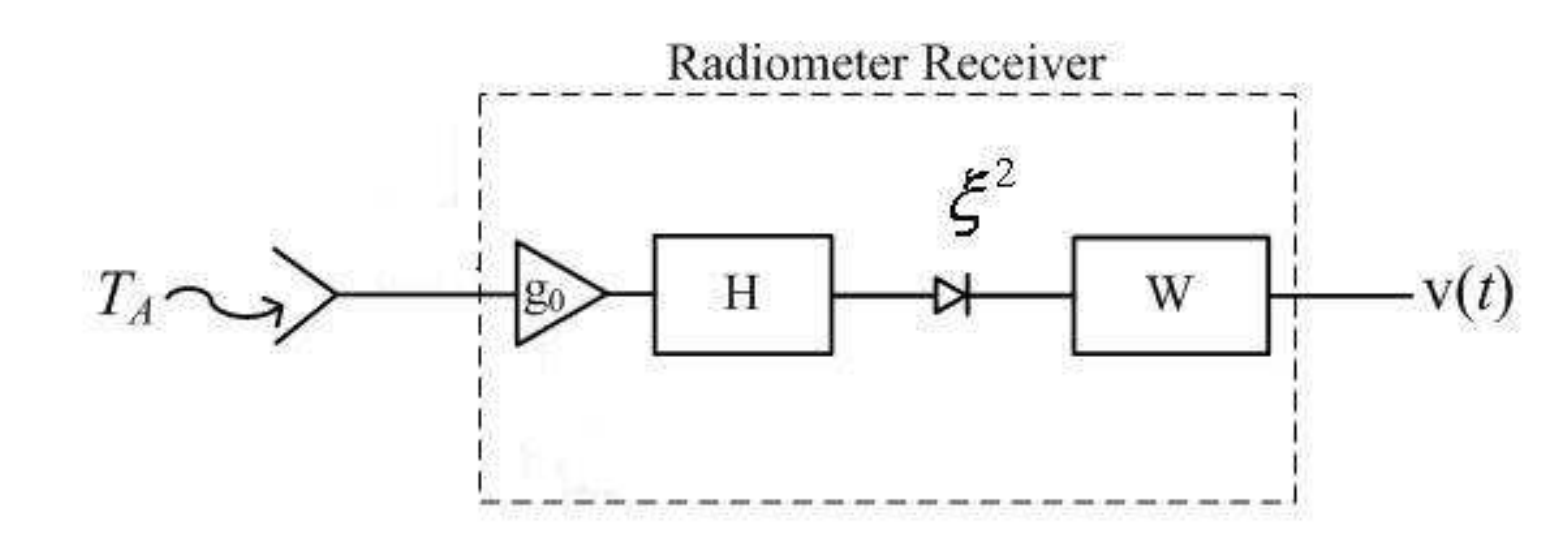}
\caption{Schematic of Simple Radiometer}
\label{fig:radiometer}
\end{figure}

\indent In this example, we apply the dynamic calibration approaches to the calibration of a microwave radiometer for an earth observing satellite. Engineers and scientist commonly use microwave radiometers to measure the electromagnetic radiation emitted by some source or a particular surface such as ice or land surface. Radiometers are very sensitive instruments that are capable of measuring extremely low levels of radiation. The transmission source of the radiant power is the target of the radiometers antenna. When the region of interest, such as terrain, is observed by a microwave radiometer, the radiation received by the antenna is partly due to self-emission by the area of interest and partly due to the reflected radiation originating from the surroundings (Ulaby {\it et al.} 1981) such as cosmic background radiation, ocean surface, or a heated surface used for the purpose of calibration.\\
\indent A basic diagram of a radiometer is shown in Figure \ref{fig:radiometer} where the radiant power with equivalent brightness temperature (i.e. the term brightness temperature represents the intensity of the radiation emitted by the scene under observation) $T_{A}$ enters the radiometer receiver and is converted to the output signal $v(t)$. 
The schematic features the common components of most microwave radiometers. As the radiometer captures a signal (i.e. Brightness Temperature $T_{A}$), it couples the signal into a transmission line which then carries the signal to and from the various elements of the circuit. In Figure \ref{fig:radiometer}, a signal $T_{A}$ is introduced directly into the antenna, then it is mixed, amplified and filtered to produce the output signal $v(t)$. This filtering and amplification of the signal is carried out through the following components of the radiometer: an amplifier $(g_{0})$; pre-detection filter $(H)$; a square law detector $(\xi^{2})$; and a post-detection filter $(W)$. The output of the radiometer is denoted as $v(t)$. See Ulaby {\it et al.} (1981) for a detailed discussion.\\	
\indent  Racette and Lang (2005) state that at the core of every radiometer measurement is a calibrated receiver. Calibration is required due to the fact that the current electronic hardware is unable to maintain a stable input/output relationship. For space observing instruments, stable calibration without any drifts is a key to detect proper trends of climate (Imaoka {\it et al.} 2010). Due to problems such as amplifier gain instability and exterior temperature variations of critical components that may cause this relationship to drift over time (Bremer 1979). During the calibration process, the radiometer receiver measures the voltage output power $v(t)$, and its corresponding input temperature of a known reference. Two or more known reference temperatures are needed for calibration of a radiometer. Ulaby {\it et al.} (1981); Racette and Lang (2005) state that the relationship between the output, $v(t)$ and the input, $T_{A}$ is approximately linear, and can be expressed as	
\begin{equation*}
\hat{T}_{A} = \beta_{0} + \beta_{1}v(t) \label{eqn:calibequation} 
\end{equation*}
where, $\hat{T}_{A}$ is the estimated value of the brightness temperature, $v(t)$ is the observed output voltage. Using this relationship, the output value, $v(t)$, is used to derive an estimate for the input, $T_{A}$ (Racette and Lang, 2005).\\
\begin{figure}[!th]
\centering
\includegraphics[width = 9cm, height = 5cm]{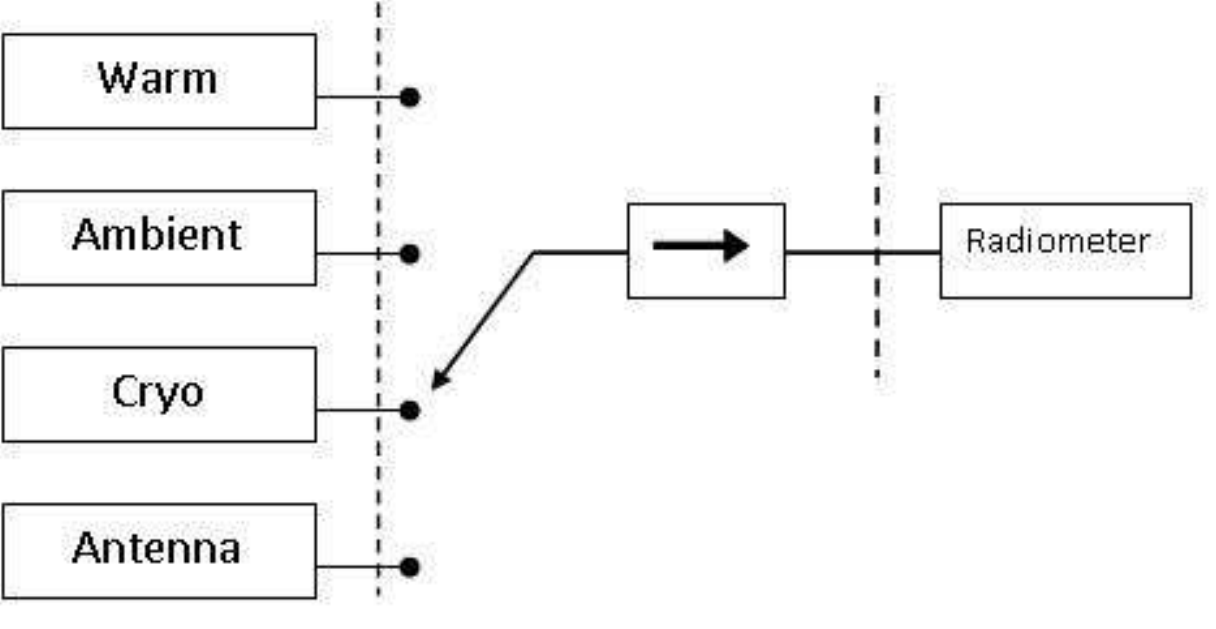}
\caption{Known Reference Temperature Collection}
\label{fig:RefData}
\end{figure} 
\indent Traditional calibration methods use measurements taken from known calibration references, for example see Figure \ref{fig:RefData}. Due to possible cost constraints it is common to use between two and five references. The reference temperatures are converted to their equivalent power measurement prior to the calibration algorithm. The radiometer outputs are observed when the radiometer measures the reference temperatures, giving an ordered calibration pair $(T_{i}, v_{i})$. The $v_{i}$ values are observed from the process of the electronics within the radiometer (see Figure \ref{fig:radiometer}) (Ulaby {\it et al.} 1981; Racette and Lang 2005). Through the process of calibration, the unknown brightness temperature $T_{j}$ is estimated by plugging its observed output $v_{j}$ into either Equation (\ref{eq:ClassEq}) or Equation (\ref{eq:InvEq}).\\ 
\indent It is of interest to develop a calibration approach that can detect gain abnormalities, and/or correct for slow drifts that affect the quality of the instrument measurements. To demonstrate the dynamic approach in terms of application appeal, the two dynamic methods were used to characterize a calibration target over time for a microwave radiometer. The data used for this example was collected during a calibration experiment that was conducted on the Millimeter-wave Imaging Radiometer (MIR) (Racette et al. 1995). The purpose of the experiment was to validate predictions of radiometer calibration.\\
\begin{figure}[!htb]
\begin{center}
\includegraphics[scale=0.35]{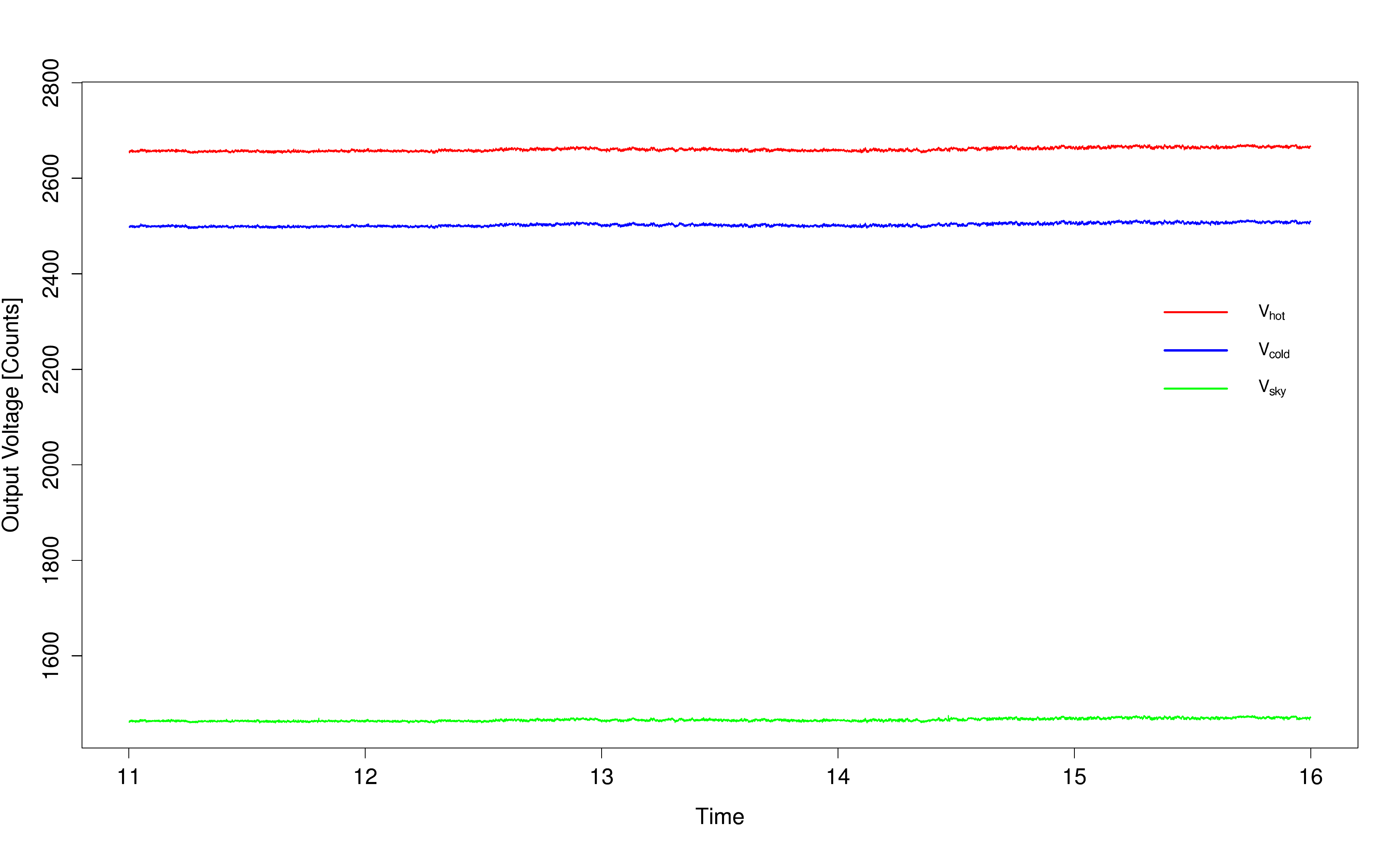}
\captionof{figure}[]{
Time series of MIR output voltage measurement data $V_{cold}$, $V_{hot}$, and $V_{sky}$.
}
\label{fig:ex_volt}
\end{center}
\end{figure}
\indent The MIR was built with two internal blackbody references which will be used to observe a third stable temperature reference for an extended period of time. The third reference was a custom designed cryogenically cooled reference. Racette (2005) conducted the MIR experiment under two scenarios: the first experiment denoted as $T295$ examined the calibration predictions when the unknown target is interior (i.e. interpolation) to the reference measurements; the second set of measurements (denoted as $T80$) where taken when the unknown temperature to be estimated is outside (i.e. extrapolation) of the range of calibration references.\\
\indent For demonstration purposes we will only consider the $T80$ experiment, for details of the $T295$ experiment see Racette (2005). For the $T80$ run of the experiment, the reference temperatures are as follows:
\begin{enumerate}
	\item $T_{cold} \sim 293.69K$	
	\item $T_{hot} ~ \sim 325.59K$
\end{enumerate}
with the unknown target temperature that must be estimated denoted as $T_{sky}$.
Each temperature measure has a corresponding observed time series of output measurements; $V_{cold}$, $V_{hot}$, and $V_{sky}$ (see Figure \ref{fig:ex_volt}). Therefore in this example we only consider a 2-point calibration set-up as we use $T_{cold}$ and $T_{hot}$ as the known reference standards and use $V_{sky}$ to derive estimates of $T_{sky}$ for the first 1000 time periods.\\
\indent The results of the dynamic approaches: $M_{D1}$ and $M_{D2}$, will be compared to the ``inverse" calibration method (Krutchkoff 1967) implemented by Racette (2005). The method considered by Racette (2005) will be denoted as $M_{1u}$. As in practice, rarely does one know the value of the true temperature to be estimated so the aim of this example is to assess the contribution of the calibration approach to the variability in the measurement estimate. Racette (2005) analysis did not consider biases that may exist in calibration, continuing in the same spirit, the existence of biases will not be considered in the analysis. We will apply the $M_{1u}$, $M_{D1}$, and $M_{D2}$ approaches to the data to estimate the temperature $T_{sky}$; the standard deviation of the estimated time series $\hat{\sigma}_{T_{sky}}$ is used as a measure of uncertainty including the contribution of the calibration algorithm.\\
\begin{figure}[!htb]
\begin{center}
\includegraphics[scale=0.35]{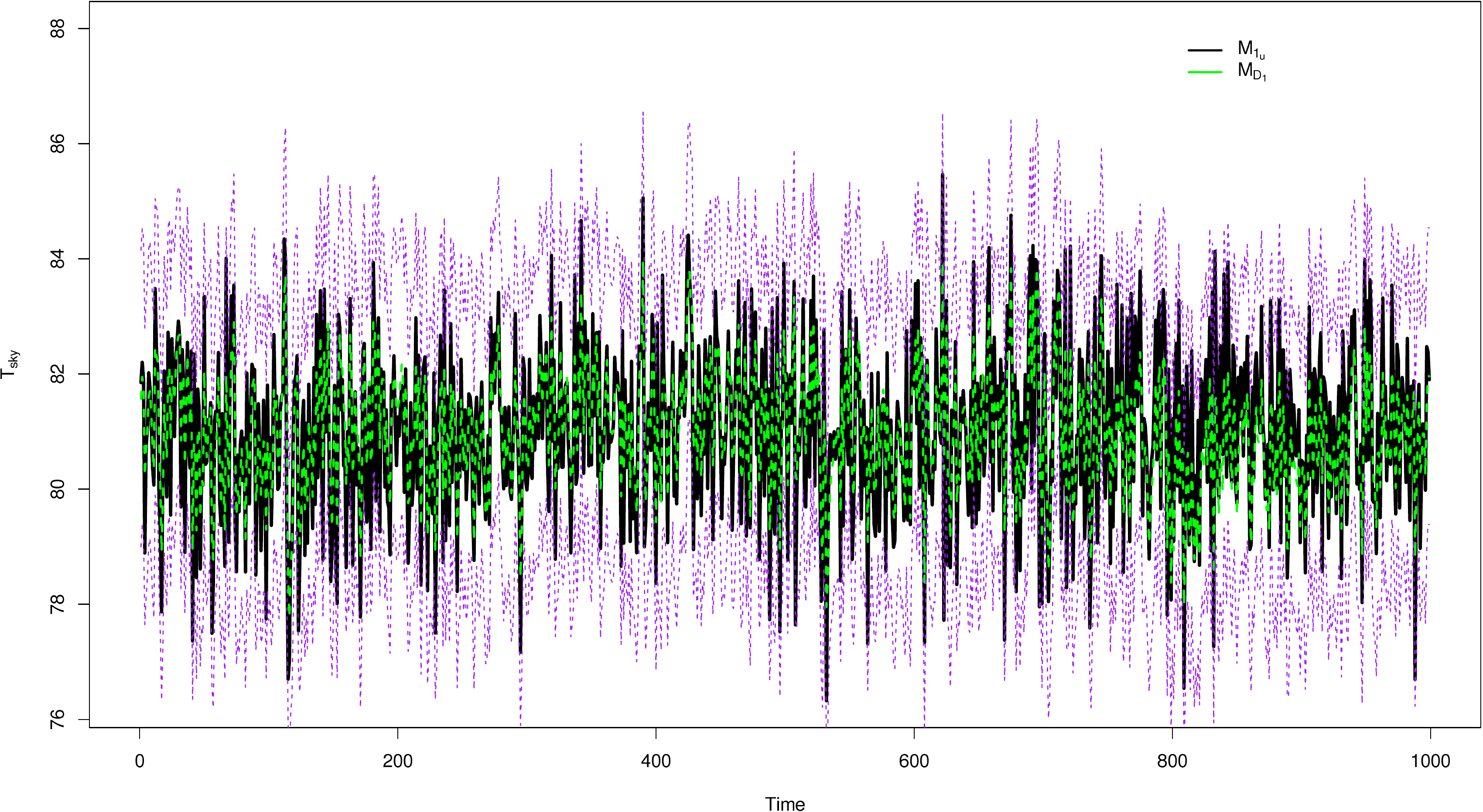}
\captionof{figure}[]{
Time series of calibrated temperature for MIR $T80$ experiment. Black lines indicate the results using the ``inverse" calibration approach $M_{1u}$; the green lines are the results using the dynamic approach $M_{D1}$ with the 95\% credible intervals in purple. 
}
\label{fig:Ex_Mod1}
\end{center}
\end{figure}
\begin{figure}[!htb]
\begin{center}
\includegraphics[scale=0.35]{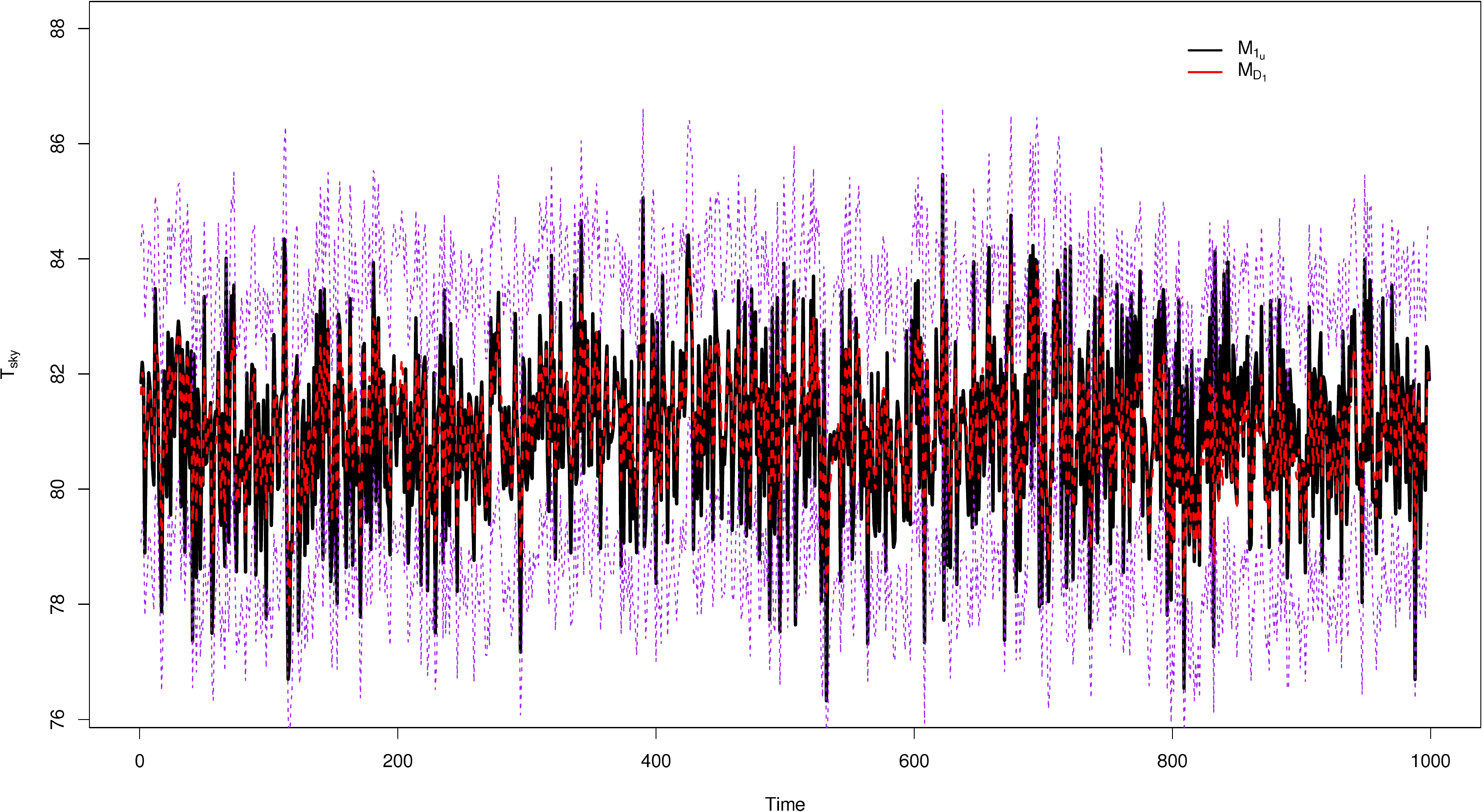}
\captionof{figure}[]{
Time series of calibrated temperature for MIR $T80$ experiment. Black lines indicate the results using the ``inverse" calibration approach $T_{1u}$; the red lines are the results using the dynamic approach $M_{D2}$ with the 95\% credible intervals in purple. 
}
\label{fig:Ex_Mod2}
\end{center}
\end{figure}
\indent Figures \ref{fig:Ex_Mod1} shows the time series of the temperature estimates for $T_{sky}$ using Krutchkoff's (1967) ``inverse" approach $M_{1u}$ and the dynamic approach $M_{D1}$. The standard deviations for the $M_{1u}$ and $M_{D1}$ approaches are $\hat{\sigma}_{T_{sky}}(M_{1u}) = 1.482K$ and  $\hat{\sigma}_{T_{sky}}(M_{D1}) = 0.998K$, respectfully. We see the dynamic model $M_{D1}$ improves the estimation process over the static model $M_{1u}$ by observing the corresponding standard deviation values. The dynamic model decreased the measurement uncertainty by roughly 33\%. In Figure \ref{fig:Ex_Mod2} the time series of the temperature estimates for $T_{sky}$ using the ``inverse" approach $M_{1u}$ and the dynamic approach $M_{D2}$ is given. The standard deviations for the $M_{1u}$ and $M_{D2}$ approaches are $\hat{\sigma}_{T_{sky}}(M_{1u}) = 1.482K$ and  $\hat{\sigma}_{T_{sky}}(M_{D1}) = 0.974K$. Again, the dynamic approach outperforms the static model $M_{1u}$. In this case, dynamic model $M_{D2}$ decreased the measurement uncertainty by roughly 34\%. 

\section{Discussion}\label{sec:future}

\begin{figure}[!htb]
\begin{center}
\includegraphics[scale=0.35]{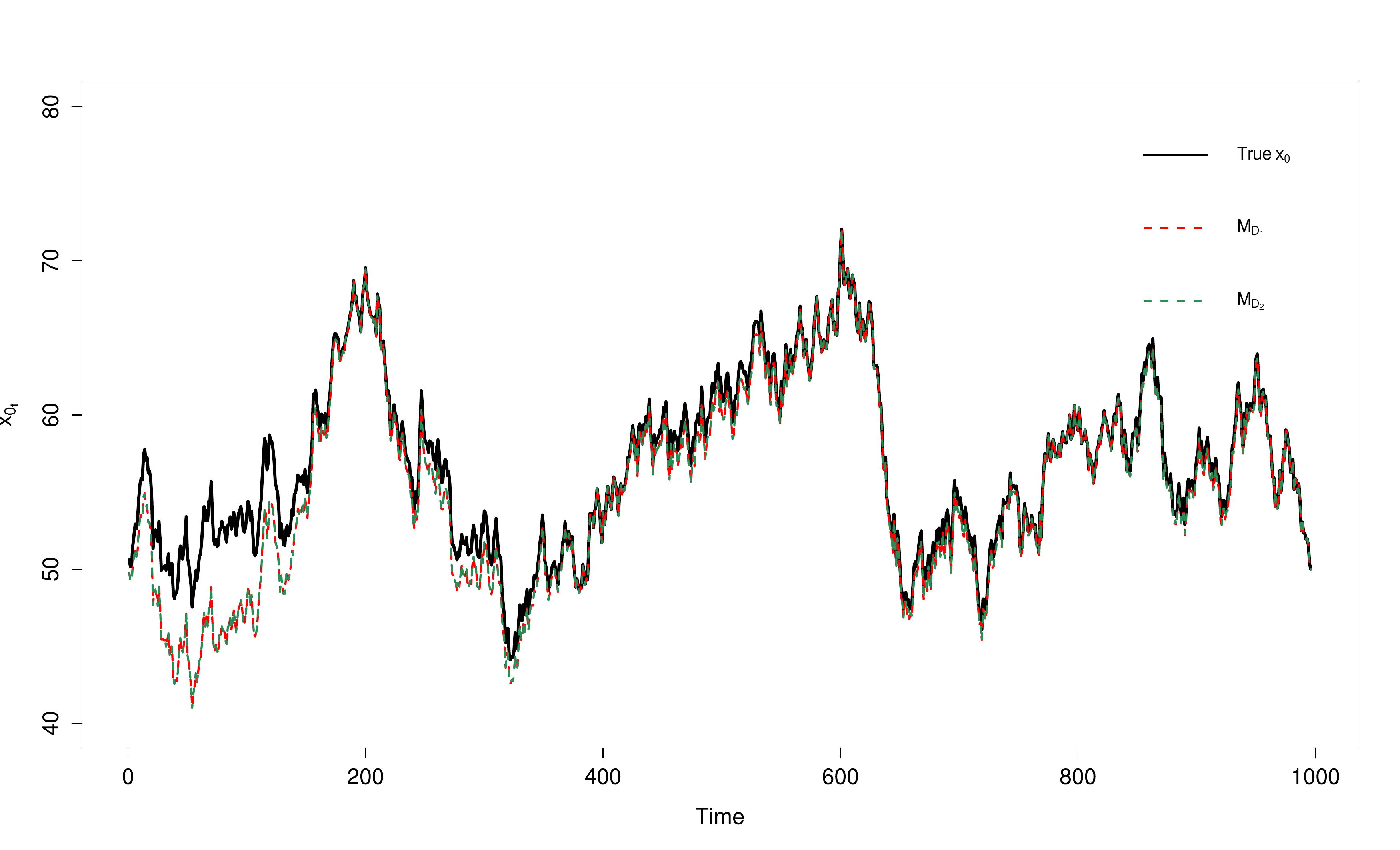}
\captionof{figure}[]{
Time series of $M_{D1}$ and $M_{D2}$ estimates in the interpolation case with gain fluctuations $g_{t}=0.1\mbox{sin}(0.025t)$ (burn-in = 0).
}
\label{fig:sine1}
\end{center}
\end{figure}

\begin{figure}[!htb]
\begin{center}
\includegraphics[scale=0.35]{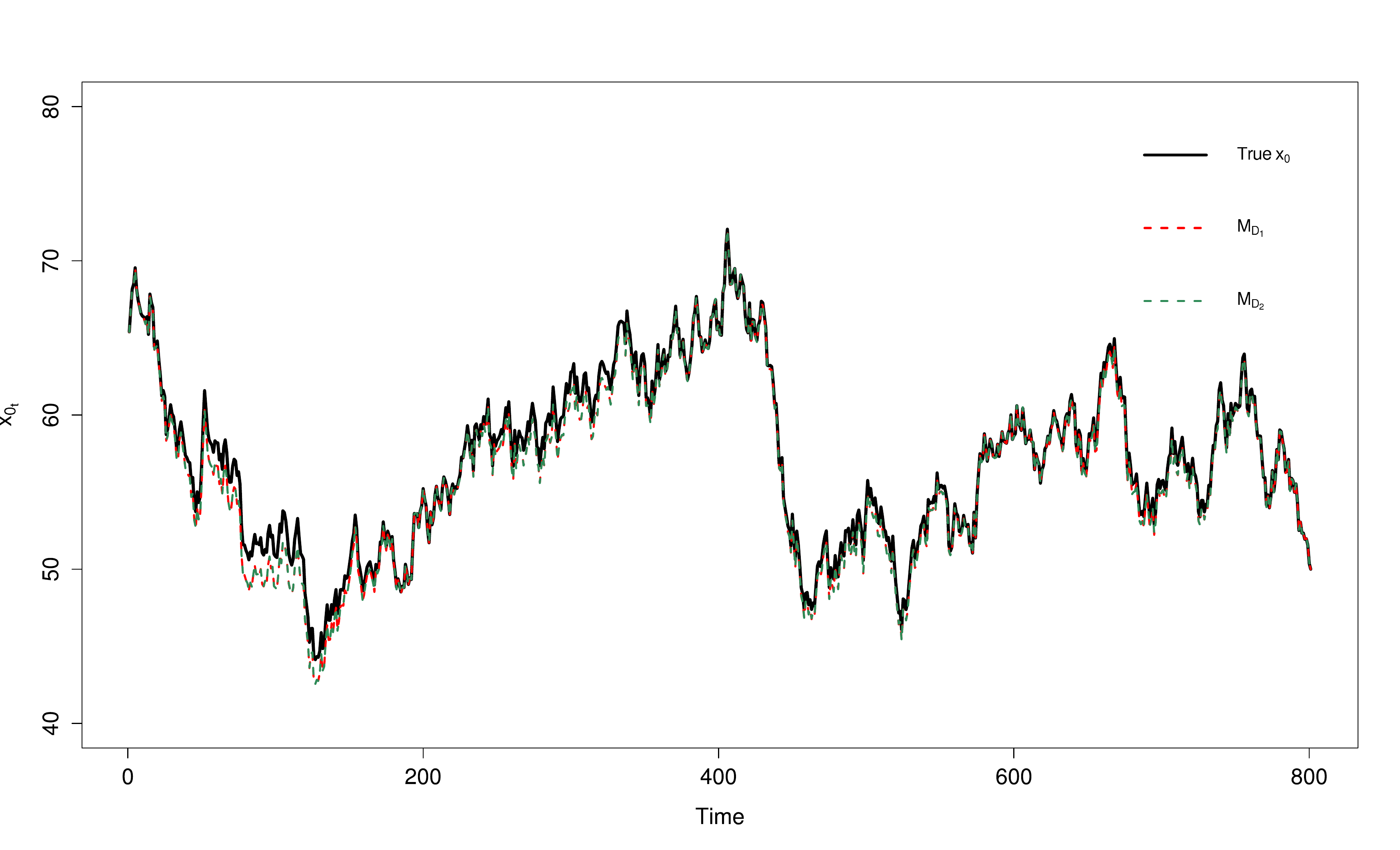}
\captionof{figure}[]{
Time series of $M_{D1}$ and $M_{D2}$ estimates in the interpolation case with gain fluctuations $g_{t}=0.1\mbox{sin}(0.025t)$ (burn-in = 200).
}
\label{tab:sine2}
\end{center}
\end{figure}

Two new novel approaches to the statistical calibration problem have been presented in this paper. In was shown by the simulation results that the use of the dynamic approach has its benefits over the static methods. If the linear relationship in the first stage of calibration is known to be stable then the traditional methods should be used. The dynamic methods showed promise in the cases when the signal-to-noise ratio was high. There is also a computation expense to implementing the dynamic methods compared to the static methods, but in the sense of electronics these methods allow for near real time calibration and monitoring. \\
\indent It is worth noting that the dynamic method shows possible deficiencies when the gain fluctuations is sinusoidal, referring to results in Table \ref{tab:sine_inter1}. In Figure \ref{fig:sine1} it is evident the largest source of the error is in the beginning of estimation process, roughly from $t=1$ to $t=200$. The MSE values for the dynamic approaches; $M_{D1}$ and $M_{D2}$ were 4.41 and 4.40, respectively, which was vastly different than those reported for the static methods. This problem can be addressed by extending the burn-in period.\\
\indent We increased the burn-in period to 200 which allowed the algorithm more time to learn and hence results in a lower MSE value. In Figure \ref{tab:sine2} we see that the estimates fit better to the true values of $x_{0t}$. The MSE decreased from 4.41 to 0.64 for $M_{D1}$ and 0.63 for $M_{D2}$. The increased burn-in period improves the coverage probability but the interval width isn't noticeably affected. The coverage probability increased from 0.628 to 0.722 for $M_{D1}$ and from 0.829 to 0.964 for $M_{D2}$.\\
\begin{table}[ht]
\centering
\captionof{table}[Comparison of calibration approaches when interpolating to estimate $x_{0t}$ with sinusoidal gain fluctuations]{
Comparison of calibration approaches $M_{D1}$ and $M_{D2}$ when interpolating to estimate $x_{0t}$ with sinusoidal gain fluctuation.
}
\begin{tabular}{c|c|c|c|}
\multicolumn{4}{c}{2 References-Sinusoidal Gain- w$/$Burn In $=200$}\\
  \toprule
 & Mean Squared Error & Coverage Probability & Interval Width  \\ 
  \hline
$M_{D1}$ & 0.63553 & 0.72185 & 1.15722 \\ 
$M_{D2}$ & 0.63333 & 0.96380 & 3.77191 \\ 
   \bottomrule
\end{tabular}
\end{table}
\begin{figure}[!htb]
\centering
\includegraphics[width = 12cm, height = 8cm]{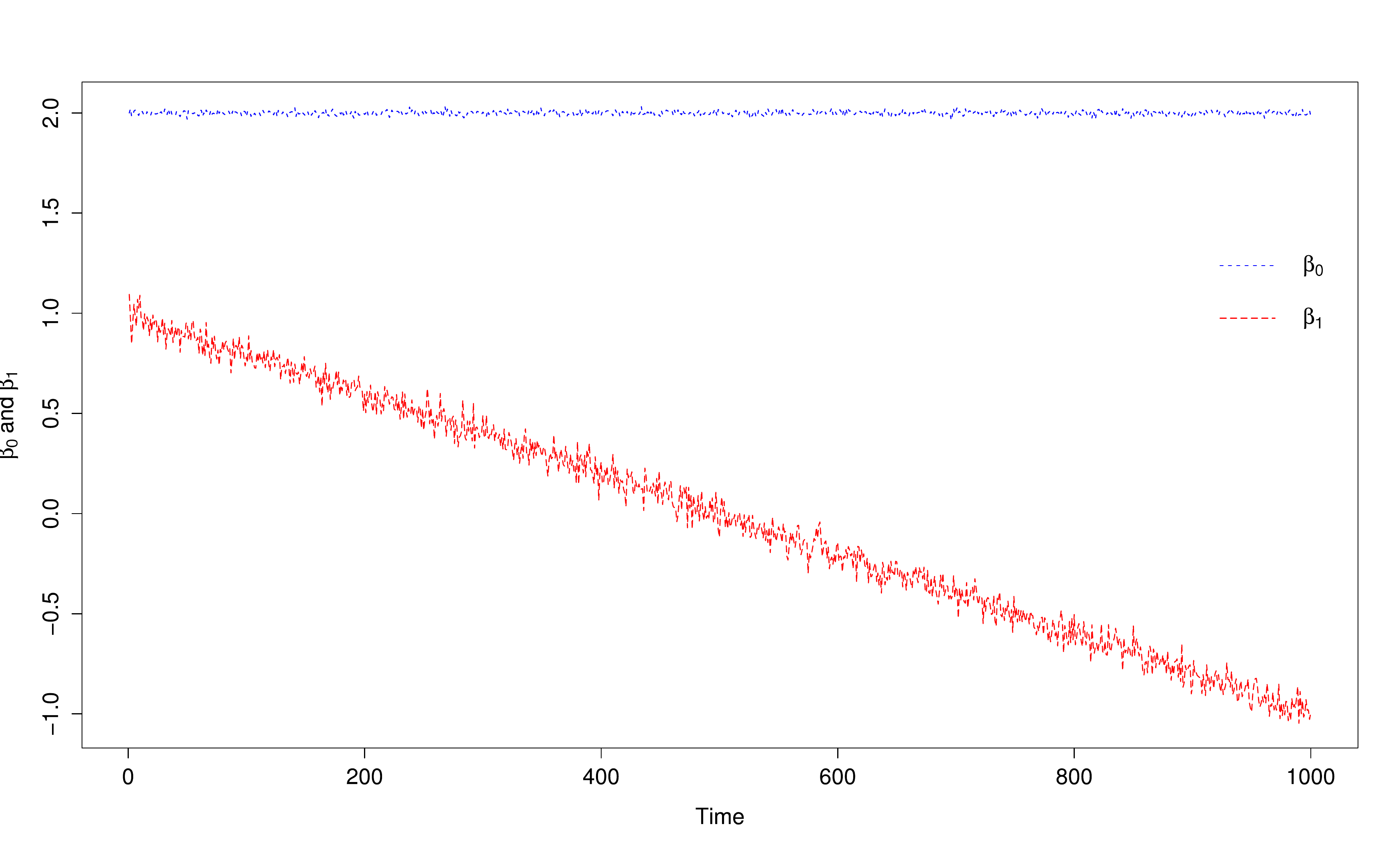}
\caption{True time series of $\beta_{0t}$ and $\beta_{1t}$}
\label{fig:beta_cross}
\end{figure}
\begin{figure}[!htb]
\centering
\includegraphics[width = 12cm, height = 8cm]{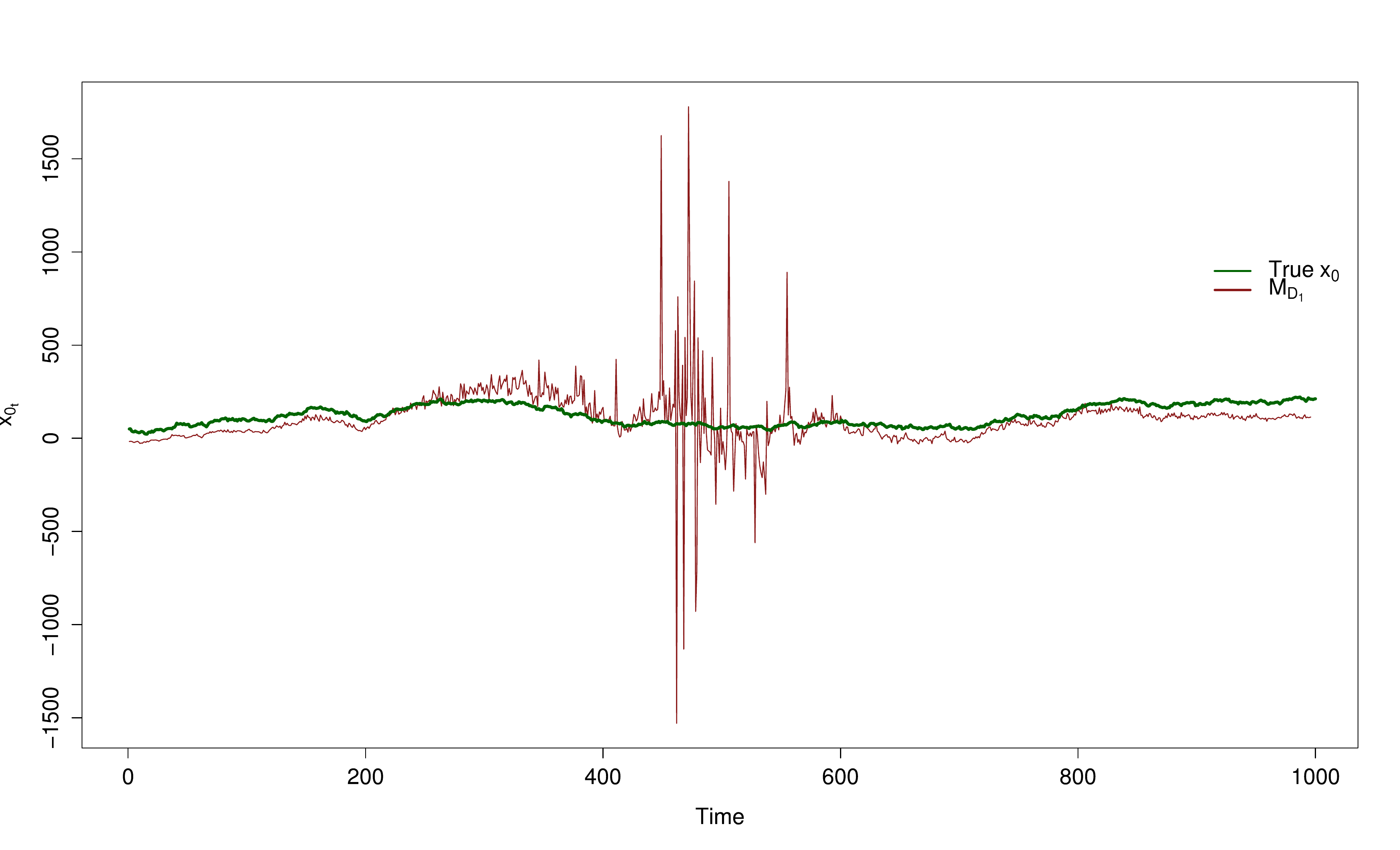}
\caption{Time series of true $x_{0t}$ and $M_{D1}$ estimate of $x_{0t}$}
\label{fig:beta_cross_x0}
\end{figure}
\indent For completeness we consider the behavior of the method if $\beta_{1t}$ crosses zero. It is absurd to believe that this would happen in practice because one would test the significance of $\beta_{1t}$ (Myers 1990; Montgomery {\it et al.} 2012) for using any method where the possibility of dividing by zero could occur. We demonstrate this by generating data where $\beta_{0t} \approx 2$ for all time and $\beta_{1t}$ drifts from 1 to -1 over time where $t=1,\dots, 1000$ (see Figure \ref{fig:beta_cross}). Figure \ref{fig:beta_cross_x0} shows the dynamic method is close to the true values of $x_{0t}$ until $\beta_{1t}$ get close to 0. Within the region where the slope crosses the $x-\mbox{axis}$ the posterior estimates become {\it unstable}. Here we define unstable as meaning that we are within a region where there is division by zero. This instability is only present when $|\beta_{1t}|<\epsilon$, for every $\epsilon > 0$. As long as $|\beta_{1t}|>0$ the dynamic method will perform well when estimating $x_{0t}$.\\ 
\indent Some calibration problems are not linear or approximately linearly related in $x_{0t}$ and $y_{0t}$. Future work is to investigate the dynamic calibration methods in the presence of nonlinearity. In such settings we may not have the ability to use only 2-points as references. Any approach will require more references in order to accurately capture the nonlinear behavior. Another area to be explored is using semiparametric regression which also allow for parameter variation across time and could be implemented in a near real time setting. 
%
%
%
%
%




\end{document}